\documentclass[english]{llncs}
\usepackage[linesnumbered,boxed, noend, noline]{algorithm2e}
\usepackage[T1]{fontenc}
\usepackage[latin9]{inputenc}
\usepackage{color}
\usepackage{float}
\usepackage{graphicx}
\usepackage{amssymb}
\usepackage{epstopdf}
\usepackage{amsmath}
\usepackage{amscd}
\usepackage{wrapfig}
\usepackage{lipsum}
\usepackage{setspace}
\makeatletter

\providecommand{\tabularnewline}{\\}
\floatstyle{ruled}
\newfloat{algorithm}{tbp}{loa}
\floatname{algorithm}{Algorithm}

\makeatother

\usepackage{babel}

\begin{document}

\def\Aut{ { \mathcal{A} } }
\def\AsupLambda{ {A^\Lambda} }
\def\AUT{ { {\it PDA} } }
\def\BAut{ {\mathcal {B} } }
\def\birwExtLambdaA{ { \smsp \buildrel {\Ext {\Lambda} (A)} \over {\longleftrightarrow} \smsp } }
\def\birwExtLambdaAstar{ { \smsp \buildrel {\Ext {\Lambda} (A)^*} \over {\longleftrightarrow} \smsp } }
\def\Kfunc{ {\it K} }
\def\birwSstarLambda{ { \smsp \buildrel {{S}^* \smsp \Lambda} \over {\longleftrightarrow} \smsp } }
\def\birwR{ { \smsp \buildrel {R} \over {\longleftrightarrow} \smsp } }
\def\birwRLambda{ { \smsp \buildrel {R \c \Lambda} \over {\longleftrightarrow} \smsp } }
\def\birwRjistar{ { \smsp \buildrel {{R_j^i}^* } \over {\longleftrightarrow} \smsp } }
\def\birwRjistarLambdai{ { \smsp \buildrel {{R_j^i}^* \smsp \Lambda_i} \over {\longleftrightarrow} \smsp } }
\def\birwRinfty{ { \smsp \buildrel {R_\infty} \over {\longleftrightarrow} \smsp } }
\def\birwRinftystar{ \smsp {\buildrel {R_\infty^*} \over \longleftrightarrow} \smsp }
\def\birwRinftyLambda{ { \smsp \buildrel {R_\infty \c \Lambda} \over {\longleftrightarrow} \smsp } }
\def\birwRiLambda{ { \smsp \buildrel {R_i \c \Lambda} \over {\longleftrightarrow} \smsp } }
\def\birwRirhoLambda{ { \smsp \buildrel {R_i \cup \{ \rho \}  \Lambda} \over {\longleftrightarrow} \smsp } }
\def\birwRinftystarLambda{ { \smsp \buildrel {R_\infty^*  \Lambda} \over {\longleftrightarrow} \smsp } }
\def\birwphistar{ { \smsp \buildrel {\emptyset^*} \over {\longleftrightarrow} \smsp } }
\def\birwphistarLambda{ { \smsp \buildrel {\emptyset^*  \Lambda} \over {\longleftrightarrow} \smsp } }
\def\birwRij{ { \smsp \buildrel {R_{i_j}} \over {\longleftrightarrow} \smsp } }
\def\birwRil{ { \smsp \buildrel { {R_{l}^{i}}^*} \over {\longleftrightarrow} \smsp } }
\def\notbirwRij{ { \smsp \buildrel {R_{i_j}} \over {\not\longleftrightarrow} \smsp } }
\def\birwRprimestar{ \smsp {\buildrel R'^* \over \longleftrightarrow} \smsp }
\def\birwRstar{ { \smsp \buildrel {R^*} \over {\longleftrightarrow} \smsp } }
\def\birwRistar{ { \smsp \buildrel {R_i^*} \over {\longleftrightarrow} \smsp } }
\def\notbirwRistar{ { \smsp \buildrel {R_i^*} \over {\not\longleftrightarrow} \smsp } }
\def\birwSstar{ \smsp {\buildrel {S^*} \over \longleftrightarrow} \smsp }
\def\birwRstarLambda{ { \smsp \buildrel {R^* \Lambda} \over {\longleftrightarrow} \smsp } }
\def\birwCR{ \smsp {\buildrel {C(R)} \over \longleftrightarrow} \smsp }
\def\birwCRstar{ \smsp {\buildrel {C(R)^*} \over \longleftrightarrow} \smsp }
\def\Bool{ {\mathbb B} }
\def\c{ , \smsp }
\def\CatAUT{ {\bf PDA}  }
\def\CatAUTplus{ {\bf PDA}^+  }
\def\CatDFA{ {\bf DFA}  }
\def\CatDA{ {\bf MA}  }
\def\child{ {\it child} }
\def\Con{ {\it Con} }
\def\concat{ {\smsp . \smsp } }
\def\conj{ \smsp \wedge \smsp }
\def\Consistent{ {\smsp {\it Cons} \smsp } }
\def\defined{ {\downarrow} }
\def\disjointsum{ {\oplus } }
\def\Dorder{ {D} }
\def\undefined{ {\uparrow} }
\def\DA{ { {\it MA} } }
\def\DFA{ { {\it DFA} } }
\def\disc{ {\it disc} }
\def\Disc{ {\it Disc} }
\def\discequiv{ {\equiv} }
\def\disj{ \smsp \vee \smsp }
\def\DSC#1#2{ {\it DSC} ( #1 \c #2 ) }
\def\EID#1#2{ { {\it E}_{#1}^{ID} {(#2)}} } 
\def\EIID#1#2{ { {\it E}_{#1}^{} {(#2)}} } 
\def\equivQ{ {\equiv_Q} }
\def\equivOmega{ {\equiv_\Omega} }
\def\endef{\vskip 6pt}
\def\endproof{\vskip 6pt}
\def\enum{ {\it enum} }
\def\eval{ {\it eval} }
\def\Ext#1{ {\it CGE_{#1} } }
\def\ff{ {\it false}}
\def\final{ {\bf 1} }
\def\FinDet{ {\it FinDet} }
\def\Future{ {\bf F} }
\def\Futurei{ {\bf F}_{\leq i} \smsp }
\def\Global{ {\bf G} }
\def\Globali{ {\bf G}_{\leq i} \smsp }
\def\Gr{ {\it Gr} }
\def\head{ {\it head } }
\def\height{ {\it height} }
\def\I#1#2{ { {\it I}_{{#1}, {#2} } } }
\def\ID{ {\it ID } }
\def\IID{ { FID } }
\def\IDS{{\it IDS}}
\def\iso{ {\smsp \cong \smsp } }
\def\inputs{\iota}
\def\Int{ {\mathbb Z} }
\def\kleeneq{ { \smsp \simeq \smsp } }
\def\Lambdaplusleqd{ { \Lambda_\ltd^+} }
\def\Lang{  {\mathfrak L}}
\def\lbrc{ \{ \sp }
\def\lbrk{ { \langle \smsp } }
\def\learn{{L}}
\def\lengthLambda{ {\vert \Lambda \vert} }
\def\leqd{ {\smsp \leq^\Dorder \smsp } }
\def\ltd{ {\smsp <^\Dorder \smsp} }
\def\lteql{ {\leq l} }
\def\max{ {\it \it max} }
\def\Max{ { {\it Max}_\prefix } }
\def\Min{ {\it Min} }
\def\mult{ \times }
\def\Nat{ {\mathbb N}}
\def\nat{ {\it nat}}
\def\Next{ {\bf X}}
\def\Nextw{ {\bf X}}
\def\Nexts{ {\bf X_s}}
\def\nf { {\it nf} }
\def\norm{ {\it norm} }
\def\notrwRstar{ \smsp {\buildrel {R^*} \over {\not \longrightarrow} } \smsp }
\def\notbirwRstar{ \smsp {\buildrel {R^*} \over {\not \longleftrightarrow} } \smsp }
\def\OneQ{ {\bf 1^Q} }
\def\OneOmega{ {\bf 1^\Omega} }
\def\output{ {\lambda} }
\def\pf{ {\sp \rightsquigarrow \sp } }
\def\PID{  {\it P}^{ID}  }
\def\PIDprime{  {\it P'}^{}  }
\def\PIID#1{  {\it P}_{#1}^{} } 
\def\PIIDprime#1{  {\it P'}_{#1}^{} } 
\def\powerset{ {\wp} }
\def\PM{ {\mathbb P} }
\def\prefix{ {\smsp \preceq \smsp } }
\def\Pref{  {\it Pref} }
\def\proj{  {\it proj} }
\def\notprefix{ {\smsp \not\preceq \smsp } }
\def\phiQ{ {\phi_{{\hskip 1pt}_Q}} }
\def\phiOmega{ {\phi_{{\hskip 1pt}_\Omega}} }
\def\rbrc{ \sp \} }
\def\rbrk{ { \smsp \rangle} }
\def\Rcupsigmasigmaprime{ { R \cup \{ \sigma \to \sigma' \} } }
\def\rhoone{ { \rho_1} }
\def\rhotwo{ { \rho_2} }
\def\rwCR{ \smsp {\buildrel C(R) \over \longrightarrow} \smsp }
\def\rwCnR{ \smsp {\buildrel C_n(R) \over \longrightarrow} \smsp }
\def\rwCnplusoneR{ \smsp {\buildrel C_{n+1}(R) \over \longrightarrow} \smsp }
\def\rwCnplusoneRstar{ \smsp {\buildrel { C_{n+1}(R)^* } \over \longrightarrow} \smsp }
\def\rwrho{ {\smsp {\buildrel \rho \over \longrightarrow} \smsp} }
\def\rwrhoi{ \smsp {\buildrel {\rho_i} \over \longrightarrow} \smsp }
\def\rwrhozero{ \smsp {\buildrel {\rho_0} \over \longrightarrow} \smsp }
\def\rwrhokminusone{ \smsp {\buildrel {\rho_{k-1}} \over \longrightarrow} \smsp }
\def\rwrhoone{ \smsp {\buildrel {\rho_1} \over \longrightarrow} \smsp }
\def\rwrhotwo{ \smsp {\buildrel {\rho_2} \over \longrightarrow} \smsp }
\def\rwR{ \smsp {\buildrel R \over \longrightarrow} \smsp }
\def\rwRi{ \smsp {\buildrel {R_i} \over \longrightarrow} \smsp }
\def\rwRirho{ \smsp {\buildrel {R_i \cup \{ \rho \} } \over \longrightarrow} \smsp }
\def\rwRprimestar{ \smsp {\buildrel R'^* \over \longrightarrow} \smsp }
\def\rwrhoprimeone{ \smsp {\buildrel {\rho'_1} \over \longrightarrow} \smsp }
\def\rwrhodoubleprimeone{ \smsp {\buildrel {\rho''_1} \over \longrightarrow} \smsp }
\def\rwrhoprimetwo{ \smsp {\buildrel {\rho'_2} \over \longrightarrow} \smsp }
\def\rwrhodoubleprimetwo{ \smsp {\buildrel {\rho''_2} \over \longrightarrow} \smsp }
\def\rwrhoprimem{ \smsp {\buildrel {\rho'_m} \over \longrightarrow} \smsp }
\def\rwrhodoubleprimen{ \smsp {\buildrel {\rho''_n} \over \longrightarrow} \smsp }
\def\rwrhostar{ \smsp {\buildrel {\rho^*} \over \longrightarrow} \smsp }
\def\rwRstar{ \smsp {\buildrel {R^*} \over \longrightarrow} \smsp }
\def\rwSstar{ \smsp {\buildrel {S^*} \over \longrightarrow} \smsp }
\def\rwCRstar{ \smsp {\buildrel {C(R)^*} \over \longrightarrow} \smsp }
\def\rwRinftystar{ \smsp {\buildrel {R_\infty^*} \over \longrightarrow} \smsp }
\def\Req{ {\it Req} }
\def\Run{ {\it Run} }
\def\RunF#1#2{ {\it RunF} ( #1 \c #2 ) }
\def\Sat{ {\it Sat} }
\def\scon{ {\it SCon} }
\def\SF{ {\it F} }
\def\SigRing#1{ {\Sigma_{#1}^{\it ring} } }
\def\sigmawave{ {\~\sigma} }
\def\sp{\hskip 5pt}
\def\smsp{\hskip 3pt}
\def\st{ \sp \vert \sp }
\def\state{ {\it state } }
\def\StateF#1#2{ {\it StateF} ( \smsp #1 \c #2 \smsp ) }
\def\startdef{\vskip 6pt}
\def\startproof{\vskip 6pt}
\def\strictprefix{ {\smsp \prec \smsp } }
\def\subs{ \leq_s }
\def\subw{ \leq_w }
\def\subtree{ {\it subtree} }
\def\suffix{ {\it suffix} }
\def\tail{ {\it tail } }
\def\TID{  {\it T}^{\ID} } 
\def\TIID#1{  {\it T}_{#1}^{} } 
\def\tower{ {T} }
\def\leqtowerd{ {\smsp \leq^{T(\Dorder)} \smsp } }
\def\lttowerd{ {\smsp <^{T(\Dorder)} \smsp } }
\def\tt{ {\it true}}
\def\twid{ {\approx} }
\def\val{ {\it out} }
\def\wcon{ {\it WCon} }

\title{An Efficient Model Inference Algorithm\\ for Learning-based Testing of Reactive Systems}
\author{Muddassar A. Sindhu}
\date{}                                           
\institute{
Computer Science Department,\\
Quaid i Azam University, Islamabad 45320, Pakistan,\\
masindhu@qau.edu.pk
}

\maketitle
\begin{abstract}
Learning-based testing (LBT) is an emerging methodology to automate iterative black-box requirements testing of software systems.
The methodology involves combining model inference with model checking techniques. However, a variety 
of optimisations on model inference are necessary in order to achieve scalable testing for large systems.

In this paper we describe the IKL learning algorithm which is an active incremental learning algorithm for deterministic Kripke structures. 
We formally prove the correctness of IKL.
We discuss the optimisations it incorporates to achieve scalability of testing. We also evaluate a black box heuristic for test termination
based on convergence of IKL learning.
\end{abstract}

\section{Introduction}
A heuristic approach to automated test case generation (ATCG) from formal requirements specifications known as 
{\it learning-based testing} (LBT) was introduced in \cite{Meinke04}, \cite{MeinkeNiu}
and \cite{MeinkeSindhu11}. 
Learning-based testing
is an iterative approach to automate specification-based black-box testing. It encompasses both test case generation, execution and evaluation (the oracle step). 
The aim of LBT is to automatically generate a large number of 
high-quality test cases by combining a model checking algorithm with an optimised model inference algorithm (aka. learning algorithm). 
For both procedural (\cite{MeinkeNiu}) and reactive systems (\cite{MeinkeNiu11}, \cite{MeinkeSindhu11}) 
it has been shown that LBT can significantly outperform random testing in the speed with which it finds errors in a system under test (SUT). 
This is because random test suites generally contain a large degree of redundancy, which can be reduced by using learning algorithms and model checkers to execute a more directed search for software errors.

An efficient and practical implementation of learning-based testing for reactive systems has been developed in the LBTest tool
\cite{MeinkeSindhu13}. In this paper we describe the IKL (Incremental Kripke Learning) algorithm implemented in LBTest. 
IKL is an algorithm for {\it active incremental learning of deterministic Kripke structures}.
The reliability of LBTest for producing correct test results depends crucially on the correctness of this learning algorithm. So we give a formal definition of  IKL and prove its correctness.
The IKL algorithm involves
a number of optimisations necessary to achieve scalability of testing for large software systems. 
We discuss these optimisations from the perspective of learning and testing.

The problems of coverage, and termination criteria for black-box testing, are complex and different solutions have been proposed. 
In LBT, convergence of learning can sometimes be used as a criterion to terminate testing. However, heuristics are needed 
to estimate convergence in the context of black box testing. We will empirically evaluate the reliability of a simple heuristic for IKL.
 
In the remainder of Section 1, we discuss the general paradigm of LBT, and specific requirements on learning for 
efficient testing of reactive systems. In Section 2, we review some essential mathematical preliminaries. In Section 3, we present the 
architecture of the IKL learning algorithm and its main components. These three main components are 
defined and analysed in detail in 
Sections 4, 5 and 6. In Section 4, we consider a learning algorithm for families of DFA which  supports 
incremental learning and projection (to be discussed in Section 1.2). In Section 5, we consider integrating a family of DFA 
into a single Kripke structure using a subdirect product construction.   In Section 6, we consider an efficient minimisation algorithm for 
deterministic Kripke structures based on Hopcroft's DFA minimisation algorithm \cite{Hop71}. This is needed by the IKL algorithm to produce hypothesis models that can be efficiently model checked. 
In Section 7, we empirically evaluate a black box heuristic to detect convergence of IKL, that can be used as a test termination criterion. Finally, in Section 8 we draw some conclusions and suggest prospects for further research on learning and testing.

\subsection{Learning-Based Testing}

The basic LBT paradigm requires three components:
\vskip 4pt
\noindent (1) a (black-box) {\it system under test} (SUT) $S$,
\vskip 4pt
\noindent (2) a {\it formal requirements specification} $\Req$ for $S$, and 
\vskip 4pt
\noindent (3) a {\it learned model} $M$ of $S$. 
\vskip 4pt
Now (1) and (2) are common to all specification-based testing, and it is really (3) that is
distinctive. Learning-based testing is a {\it heuristic iterative method} to
automatically generate a sequence of test cases. The heuristic concept is to {\it learn a black-box system using 
tests as queries}.

In general, an LBT algorithm iterates the following four steps: 
\vskip 4pt
\noindent (Step 1) Suppose that $n$ test case inputs $i_1 \c \ldots \c i_n$ have been executed 
on $S$ yielding the system outputs $o_1 \c \ldots \c o_n$. The $n$ input/output observations
$(i_1 \c o_1) \c \ldots \c$ $(i_n \c o_n)$ can be synthesized into a learned model $M_n$
of $S$ using an {\it incremental learning algorithm} (see Section 1.2). This step involves {\it generalization} from the observed 
behaviour, (which represents an incomplete description of $S$) to all possible behaviour.
This generalisation step gives the possibility to predict previously unseen errors in
$S$ during Step 2.
\vskip 4pt
\noindent (Step 2) The system requirements $\Req$ are checked against the learned model $M_n$ derived in Step 1 (aka. {\it model checking}).
This process searches for a  {\it counterexample} $i_{n+1}$ to the requirements. 
\vskip 4pt
\noindent (Step 3) The counterexample $i_{n+1}$ is executed as the 
next  test case on $S$, and if $S$ terminates then the
output $o_{n+1}$ is obtained. 
If $S$ fails this test case (i.e. the observation $( i_{n+1} \c o_{n+1} )$ does not satisfy $\Req$)
then $i_{n+1}$ was a {\it true negative} and we proceed to Step 4. Otherwise $S$ passes the 
test case $i_{n+1}$ so the model $M_n$ was inaccurate, and $i_{n+1}$ was a {\it false negative}. 
In this latter case, the effort of executing $S$ on $i_{n+1}$ is not wasted. We return to Step 1
and apply the learning algorithm once again to $n+1$ pairs
$(i_1 \c o_1) \c \ldots \c (i_{n+1} \c o_{n+1})$ to infer a refined model $M_{n+1}$ of $S$.
\vskip 4pt
\noindent (Step 4) We terminate with a true negative test case ($i_{n+1}$, $o_{n+1}$) for $S$.
\vskip 4pt

Thus an LBT algorithm iterates Steps $1 \dots$ 3 until 
an SUT error is found (Step 4) or execution is terminated. 
Practical criteria for termination of testing include a bound on the maximum testing time, or a bound on the maximum number of
test cases to be executed. However, it also seems possible to derive more theoretically well-founded criteria for termination based on learning theory.
One simple approach will be discussed in Section 7. A more sophisticated proposal can be found in 
\cite{Walkinshaw2011}.

This iterative approach to automated test case generation yields a sequence of increasingly accurate models 
$M_0$, $M_1$, $M_2$, $\ldots$, of $S$. 
(We usually take $M_0$ to be a null hypothesis about $S$.) 
So, with increasing values of $n$, it becomes more and more likely that model checking in Step 2 will 
produce a true negative if one exists. 

Notice, if Step 2 does not produce
any counterexamples at all then to proceed with the next iteration, we must construct the next test case
$i_{n+1}$ by some other method. Now  active learning algorithms can be devised to generate queries
that efficiently learn an unknown system in polynomial time.  So for LBT there is clearly an advantage to combine model checking with active learning
and generate both types of test cases. 
More generally, it is useful to have access to as wide a variety of query generation techniques as possible.
So in practice, model checker and active learning queries are augmented with random queries when necessary. However, these different types of queries
need to be  combined carefully to achieve efficient and scalable testing.

\subsection{Learning for Efficient Testing}
As has already been suggested in Section 1.1, for LBT to be effective at finding errors, it is important to use the right kind of learning algorithm. As well as active learning, several other principles for efficient testing can be found. To motivate the design of the IKL algorithm we will discuss two of them. For this purpose, we focus specifically on automata learning for testing {\it reactive 
systems}. (LBT has also been successfully applied to testing other types of systems, see e.g. \cite{MeinkeNiu}). Learning algorithms for automata are also known as {\it regular inference algorithms}  in the literature (e.g. \cite{Higuera}).

\subsubsection{Incremental Learning}
One efficiency principle is that a good learning algorithm should maximise the opportunity of the model checker in Step 2 above to find a true counterexample $i_{n+1}$ to the requirements $\Req$ as soon as possible. 
An automata learning algorithm $L$ is said to be {\it incremental} if it can produce a sequence of hypothesis automata 
$\Aut_0 \c \Aut_1 \c \ldots $ which are approximations to an unknown automata $\Aut$, based on a sequence of observations of the input/output behaviour of $\Aut$. The sequence $\Aut_0 \c \Aut_1 \c \ldots $ must finitely converge to 
$\Aut$, at least up to behavioural equivalence. In addition, the computation of each new approximation $\Aut_{i+1}$
by $L$ should reuse as much information as possible about the previous approximation $\Aut_{i}$ (e.g. equivalences between states). 
Incremental learning algorithms are necessary for two reasons.
\vskip 4pt
\indent (1) Real world systems are often too big to be completely learned and tested within a feasible timescale. This is mainly 
due to: (i) the time complexity of learning and model checking algorithms, and (ii) the time needed to execute the 
individual test cases on a large SUT.

\indent (2) Testing of specific requirements such as use cases may not require learning and analysing the entire SUT $S$, but only  the relevant fragment of $S$ which implements the requirement $\Req$.

For these two reasons, the IKL learning algorithm used in LBTest is based on incremental learning.

This concept of a {\it relevant fragment} of an SUT for testing a requirement $\Req$ raises the question of the 
relative efficiency of different types of queries (test cases). We have already seen that in LBT, test cases can be generated by 
model checking, by active learning, or by some other process entirely such as random querying.

As indicated in (1) above, the overhead of SUT execution time to answer an individual query can be large compared with the execution time of learning and model checking. There are examples of industrial systems where this execution time is of the order 
of minutes. So realistically, queries should be seen as ``expensive''.
From the viewpoint of relevance therefore, as many queries as possible should be derived from model checking the hypothesis automaton, since these queries are all based on checking the requirements $\Req$. Conversely as few queries as possible should be derived from the active learning algorithm. 
Active learning queries have no way to reference the requirement $\Req$, and therefore can only uncover an SUT error by accident. 
Furthermore, active learning queries may explore parts of the SUT which are irrelevant to checking $\Req$, thereby 
leading the search for errors in a fruitless direction. Ideally, {\it every} query would represent a relevant and interesting requirements-based test case. 

However, there is conflicting  issue involved here, which is the computational effort needed to generate different types of queries. Model checker generated queries are generally computationally expensive relative to active learner generated queries, 
often by several orders of magnitude. Therefore, if too many (perhaps even all) queries are generated by model checking, 
then the LBT process may slow so much that random testing is simply faster. In a practical LBT tool, the ratio between
the number of model checker generated queries, and the number of active learning queries must be controlled to achieve a
balance between relevance and speed. The IKL algorithm implements a pragmatic balance between these two types of queries
that we have found to be reasonably efficient in practise. 

Interestingly, when the balance of active learning queries becomes very high,  and model checking queries are almost 
eliminated, we might think that LBT becomes similar to random testing. However 
\cite{Walkinshaw}
shows that this is not the case. Thus using active learner queries alone, LBT can achieve better functional coverage than random testing.

\subsubsection{Projection}
When we consider the output variables of the SUT $S$ that appear in a specific formal black box requirement $\Req$, we often see just a small subset of the set of all output variables of $S$. 
This observation points to a powerful abstraction technique for learning that can be termed 
{\it bit-slicing} (for propositional variables) or more generally {\it projection}. 

Like incremental learning, projection is another abstraction method 
that concentrates on learning only the relevant SUT behavior needed to test the requirement $\Req$. Essentially, projection 
involves learning a quotient model of the SUT by observing just the output variables
appearing in $\Req$. Since quotient models of $S$ may be dramatically smaller than $S$ itself, the time needed for learning and testing may be considerably reduced. 
Therefore, projection seems to be an essential component of a scalable LBT system. Indeed, the combination of incremental learning 
and projection seems to be particularly powerful.
The IKL algorithm incorporates 
both these features, and they will be discussed in further detail in Sections 3 and 4.

\subsection{Literature Survey}
Several previous works, (for example Peled et al.  \cite{Peled}, Groce et al. \cite{Groce} and Raffelt et al. \cite{Raffelt}) have considered a combination of learning and model checking to achieve testing and/or formal verification of reactive systems. 
Within the model checking community the verification approach known as 
{\it counterexample guided abstraction refinement} (CEGAR) also combines learning and model checking, (see e.g. Clarke et al. \cite{Clarke} and Chauhan et al. \cite{Chauhan}). The LBT approach can be distinguished from these other approaches by: (i) an emphasis on testing rather than verification, and (ii) the use of {\it incremental learning}  and other abstraction techniques specifically chosen to achieve scalable testing and faster error discovery (c.f. Section 1.2).

In practise, most of the well-known classical regular inference algorithms such as L* (Angluin \cite{Angluin87}) or ID 
(Angluin \cite{AngluinID}) are designed for complete rather than incremental learning. Among the much smaller number of known incremental learning algorithms, we can mention the RPNII algorithm
(Dupont \cite{Dupont96}) and the IID algorithm (Parekh et al. \cite{Parekh}) which learn Moore automata, and the ICGE algorithm 
(Meinke and Fei \cite{MeinkeNiu12}) which learns Mealy automata over abstract data types. No algorithm which combines incremental learning and projection has been published in the literature. The problem of integrating active learning queries with model checker generated 
queries (which in some sense take over the role of Angluin's equivalence checker \cite{Angluin87}) has also not been considered. 
Thus: (i) the {\it design of the IKL algorithm}, (ii) its {\it formal proof of correctness}, and (iii) its {\it motivation by efficient test case generation} represent the main novel contributions of our paper.

The use of minimisation algorithms in automata learning also seems not to have been considered. This is mainly  
because most DFA learning algorithms 
naturally infer the canonical minimal automaton. However, our use of projection as an abstraction method for learning large Kripke structures
does not lead immediately to minimal structures. In fact, inferring non-minimal automata can even lead to efficiency gains 
as we have shown elsewhere in \cite{MeinkeNiu12}. 

For different automata models and different notions of equivalence, 
 the complexity of the minimisation problem can vary considerably.  The survey \cite{Ber11} considers minimisation algorithms for DFA up to language equivalence, 
 with time complexities varying between $\mathcal O(n^{2})$ and $\mathcal O(n\;log\;n)$. 
Kripke structures represent a generalisation of DFA to allow non-determinism and multiple outputs. They
have been widely used to model concurrent and embedded systems. 
An algorithm for mimimizing Kripke structures has been given in
\cite{BusGru03}. 
In the presence of non-determinism, the complexity of minimisation is quite high. 
Minimisation up to language equivalence requires exponential time, while 
minimisation up to a weaker simulation equivalence can be carried out polynomial time
(see \cite{BusGru03}).
By contrast, we will show that {\em deterministic} Kripke structures can be efficiently minimized even up to language equivalence
with a worst case time complexity of $\mathcal O(kn \log_{2} n)$. 
Our generalisation of Hopcroft's DFA minimisation algorithm to deterministic Kripke structures in Section 6 is fairly simple and straightforward.
Nevertheless, this algorithm has not been previously published in the literature, and represents another novel contribution.

\section{Mathematical Preliminaries and Notation}
In this section we introduce some basic concepts and notations needed to define and prove the correctness of the 
IKL learning algorithm.
Let $\Sigma$ be any set of symbols then $\Sigma^{*}$ denotes
the set of all finite strings over $\Sigma$ including the empty string $\varepsilon$. The length of
a string $\alpha \in \Sigma^{*}$ is  denoted by $\vert\alpha\vert$ and $\vert \varepsilon \vert=0$. 
For strings $\alpha,\beta \in\Sigma^{*}$, $\alpha \concat \beta$ denotes their concatenation.

For $\alpha,\beta,\gamma \in\Sigma^{*}$,
if $\alpha=\beta \gamma$ then $\beta$ is termed a {\it prefix} of
$\alpha$ and $\gamma$ is termed a {\it suffix} of $\alpha$. We let
$\Pref(\alpha)$ denote the prefix closure of $\alpha$, i.e. the set of all prefixes of 
$\alpha$. We can also apply prefix closure pointwise to any set of strings. The {\it set difference operation} between two sets $U,V$,
denoted by $U - V$, is the set of all
elements of $U$ which are not members of $V$. The {\it symmetric difference}
operation on pairs of sets is defined by
$U \oplus V = (U - V) \cup (V - U)$. 

A {\it deterministic finite automaton} (DFA) is a five-tuple $\Aut = (\Sigma,Q,F,q_{0},\delta)$
where: $\Sigma$ is the input alphabet, $Q$ is the state set, $F\subseteq Q$
is the accepting state set and  $q_{0}\in Q$ is the starting state. 
The state transition function of $\Aut$ is a mapping $\delta : Q \mult \Sigma \to Q$ with the usual meaning, and can be inductively extended
to a mapping $\delta^* : Q \mult \Sigma^* \to Q$ where
$\delta^* ( q \c \varepsilon ) = q$ and 
$\delta^* ( q \c \sigma_1 \c \ldots \sigma_{n+1} ) = \delta (  \delta^* ( q \c \sigma_1 \c \ldots \sigma_{n} ) \c \sigma_{n+1} ) $.

A  {\it dead state} is a state from which no accepting state can be reached, 
and a state which is not dead is termed {\it live}.
Since input strings can be used to name states, 
given any distinguished  dead state $d_0$ we define 
 {\it string concatenation modulo the dead state} $d_0$,
$f :  \Sigma^* \cup  \{ d_0 \} \mult \Sigma \to \Sigma^* \cup  \{ d_0 \}$, by $f( d_0 \c \sigma ) = d_0$
and $f(  \alpha \c \sigma ) =  \alpha \concat \sigma$ for $\alpha \in \Sigma^*$. 
This function is used for automaton learning in Section 4. 

The {\it language} $L(\Aut)$ {\it accepted by} $\Aut$ is the set of all strings $\alpha \in \Sigma^*$ such that 
$\delta^* ( q_0 \c \alpha ) \in F$. 
A language $L \subseteq \Sigma^*$ is accepted by a DFA 
if and only if, $L$ is {\it regular}, i.e. $L$ can be defined by a regular grammar.

A generalisation of DFA to allow multi-bit outputs on states  is given by deterministic Kripke structures.

\startdef
\noindent {\bf 2.1. Definition.}
{ Let $\Sigma = \lbrc  \sigma_1  \c \ldots \c  \sigma_n \rbrc$ be a finite input alphabet. By a $k$-bit
{\it deterministic Kripke structure}  $\Aut$ we mean a five-tuple
$$
\Aut = (  \smsp Q_\Aut  \c \Sigma  \c \delta_\Aut : Q_\Aut  \mult  \Sigma  \to Q_\Aut  \c q_\Aut^0 \c  \lambda_\Aut : Q_\Aut  \to \Bool^k   \smsp )
$$
where $Q_\Aut$ is a state set, $\delta_\Aut$ is the state transition function, $q_\Aut^0$ is the initial state and 
$\lambda_\Aut$ is the output function. 

As before we let $\delta_\Aut^* : Q_\Aut  \mult  \Sigma^*  \to Q_\Aut$ denote the iterated 
state transition function, where $\delta_\Aut^* ( q \c  \varepsilon ) = q$ and 
$\delta_\Aut^* ( q \c  \sigma_1  \c \ldots \c  \sigma_{i+1} ) = 
\delta_\Aut ( \delta_\Aut^* ( q \c  \sigma_1  \c \ldots \c  \sigma_{i} ) \c \sigma_{i+1} ) $. Also we let 
$\lambda_\Aut^* :   \Sigma^*   \to \Bool^k$ denote the iterated 
output function 
$\lambda_\Aut^* ( \sigma_1  \c \ldots \c  \sigma_{i} ) = 
\lambda_\Aut (  \delta_\Aut^* ( q_\Aut^0 \c  \sigma_1  \c \ldots \c  \sigma_{i} )  )  $.
More generally for any $q  \in Q$ define 
$\lambda_{q}^{*}(\sigma_{1},...,\sigma_{i})= \lambda_\Aut (  \delta_\Aut^* ( q \c  \sigma_1  \c \ldots \c  \sigma_{i} )  ) $. 
Given any $R\subseteq Q$ we write $\lambda(R)=\cup_{r\in R}\lambda(r)$. We let $q. \sigma$ denote $\delta(q,\sigma)$ and $R. \sigma$ denotes $\{r. \sigma \;\vert\; r \in R\}$ for $R\subseteq Q$.

}
\endef
Note that a $1$-bit deterministic Kripke structure $\Aut$ 
is isomorphic to a DFA 
$\Aut' = (  \smsp Q_{\Aut'}  \c \Sigma  \c \delta_{\Aut'} : Q_{\Aut'}  \mult  \Sigma  \to Q_{\Aut'}  \c q_{\Aut'}^0 \c  F_ {\Aut'}  \smsp )$, where 
$F_ {\Aut'} \subseteq Q_{\Aut'}$ and $\lambda_{\Aut'} (q) = \tt$ if, and only if 
$q \in F_ {\Aut'}$. 

In the context of Boolean valued output variables, the concept of projection on a set of output variables will also 
be termed {\it bit slicing}. 
Let us make precise the concept of a bit-slice or projection of a Kripke structure. 
\startdef
\noindent {\bf 2.2. Definition.}
{
Let ${\Aut}$ be a $k$-bit Kripke structure over a finite input alphabet $\Sigma$,
$$
{\Aut} = (  \smsp Q_{\Aut}  \c \Sigma  \c \delta_{\Aut} : Q_{\Aut}  \mult  \Sigma  \to Q_{\Aut}  \c q_{\Aut}^0 \c  \lambda_{\Aut} : Q_{\Aut}  \to \Bool^k   \smsp ) .
$$
For each $1  \leq i \leq k$ define the {\it $i$-th projection} ${\Aut}_i$ of ${\Aut}$ to be the $1$-bit Kripke structure 
where 
$$
{\Aut}_i = (  \smsp Q_{\Aut}  \c \Sigma  \c \delta_{\Aut} : Q_{\Aut}  \mult  \Sigma  \to Q_{\Aut}  \c q_{\Aut}^0 \c  \lambda_{{\Aut}_i} : Q_{\Aut}  \to \Bool   \smsp ) ,
$$
and $\lambda_{{\Aut}_i} (q) = \lambda_{\Aut} (q)_i$, i.e. $\lambda_{{\Aut}_i} (q)$ is the $i$-th bit of $\lambda_{\Aut} (q)$.
}
\endef

A family of $k$ individual $1$-bit Kripke structures can be combined into a single $k$-bit Kripke structure using a subdirect product construction.
This will be discussed in Section 5. 

A Kripke structure ${\Aut}$ is {\it minimal} if it has no proper subalgebra. This is equivalent to all states of ${\Aut}$ being reachable 
from the initial state by means of some input string. 
If ${\Aut}$ is a Kripke structure then ${\Aut}$ always has a  {\it minimal subalgebra} which we denote by $Min({\Aut})$.

\section{Architecture of the IKL Algorithm}

\begin{figure}
  \centering
    \includegraphics[width=1\columnwidth,height=0.30\paperheight]{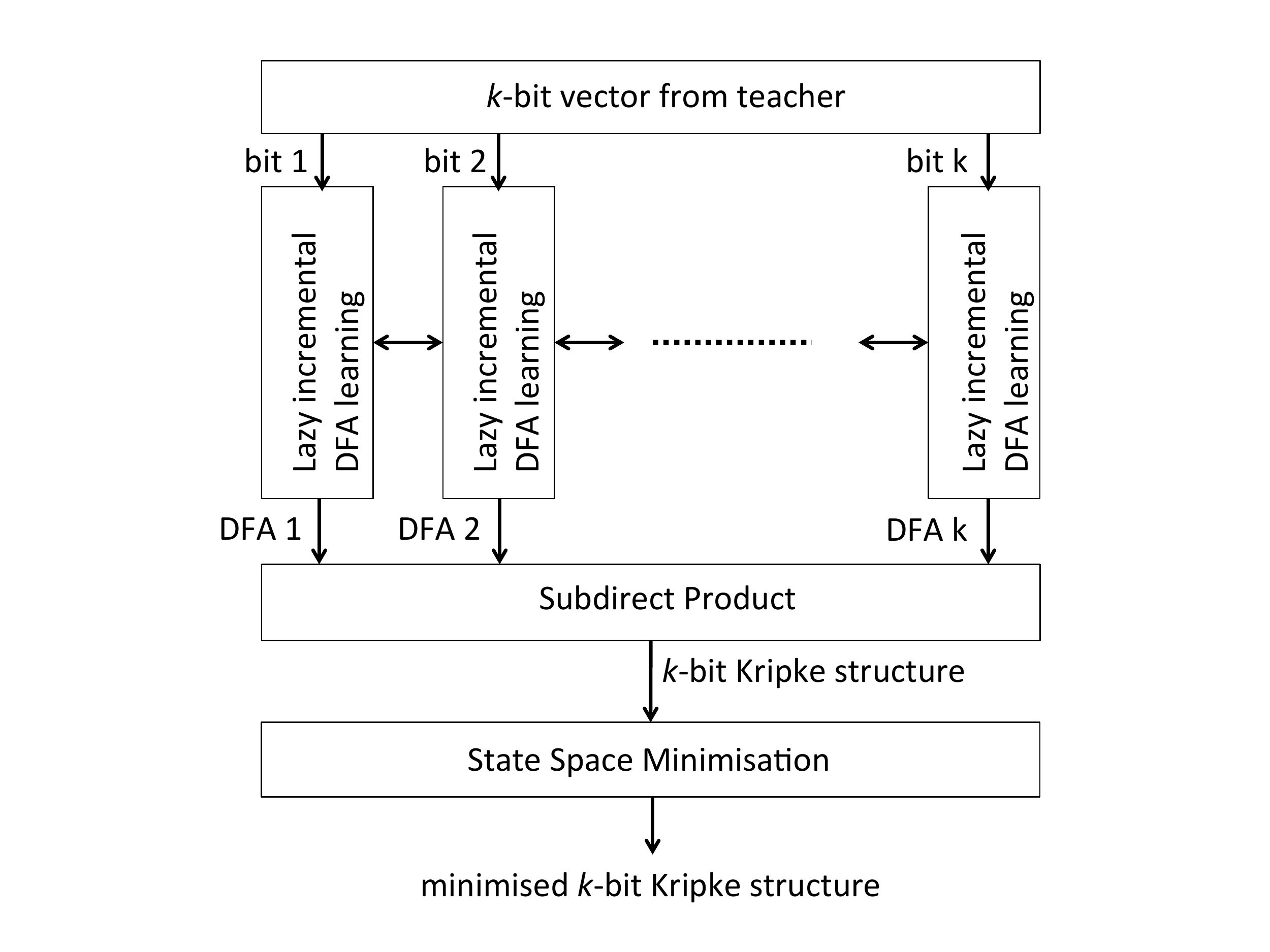}
    \caption{\label{architecture}Architecture of the IKL algorithm.}
  
\end{figure}

As discussed in Section 1,  IKL is an algorithm for incrementally 
inferring a deterministic $k$-bit Kripke structure from observational 
data.
For efficient testing, it also implements projection on output variables. An architectural view of the IKL algorithm is given in Figure \ref{architecture}. The basic idea of the algorithm 
is to learn a $k$-bit Kripke structure as a family of $k$ $1$-bit Kripke structures (i.e. DFA) using an incremental DFA learning 
algorithm for each of the $k$ individual DFA. 

For DFA learning, we use an incremental refinement of Angluin's ID algorithm \cite{AngluinID}. 
The dead state $d_0$ used in the ID algorithm can be used in incremental learning as an abstraction for all currently unknown information 
about the system to be learned.
Our refinement of ID  differs from the IID learning algorithm described in \cite{Parekh} in ways which have been discussed in \cite{MeinkeSindhu12a}. 
Note that the IKL architecture is modular, in the sense that other DFA learning algorithms could be used instead of ID.
These might alter the overall performance of IKL from a testing perspective (see Section 8).

As Figure \ref{architecture} indicates, 
these $k$ incremental DFA learning algorithms must co-operate in order to 
jointly learn an entire family of DFA. This co-operation between the DFA learners is 
termed {\it lazy learning}. It is used to  support more frequent model checking during testing, which is desirable for the reasons
explained in Section 1.2. The goal of lazy learning then is to learn the DFA family in a way that can produce new
$k$-bit hypothesis Kripke structures with maximum frequency. 

The $k$ individual DFA are assembled into a single $k$-bit Kripke structure using a generalisation  of the
direct product construction known as a {\it subdirect product}. Without minimisation, the state space of the product automaton would be very large.
However, the state space can be reduced on the fly, resulting in a subalgebra of the direct product. This removes all states 
which are not reachable from the initial state, using some input string. The state space of this subdirect product is typically 
still very large. In order to minimise the state space even further, we finally apply a minimisation algorithm for Kripke structures.
For this we adapt Hopcroft's  minimisation algorithm for DFA \cite{Hop71},  and generalise it to Kripke structures. We will discuss the state space sizes achieved by the intermediate Kripke structures during the incremental learning process in Section 7. 

In the next three sections we define and prove correct the three major
components of the IKL algorithm:  \ \

 \indent (i) the DFA family learning algorithm $\IID$ (Section  4), 
 
 \indent (ii) the subdirect product construction (Section 5), and 
 
 \indent (iii) a Kripke structure minimisation algorithm (Section 6)


\section{Incremental Learning of DFA Families}

In this section we define and prove correct an algorithm $\IID$ for incremental lazy learning 
of a family of DFA that share a common input.
This approach supports bit-sliced learning of a large Kripke structure by
projection of specific output variables (c.f. Section 1.2). Our algorithm is derived from the $\ID$ learning algorithm for DFA described 
in \cite{AngluinID}, and our correctness proof makes use of the correctness property of $\ID$. Therefore, we begin by 
reviewing the $\ID$ algorithm itself, before turning our attention to DFA family learning.

\subsection{The $\ID$ Algorithm}
The $\ID$ algorithm and  its correctness have been discussed at length  in \cite{AngluinID}. Therefore our own presentation can be brief. 
A finite set $P \subseteq \Sigma^*$ of input strings 
is said to be {\it live complete} for a DFA $\Aut$ if for every live state $q \in Q$ there exists a string 
$\alpha \in P$ such that $\delta^* ( q_0 \c \alpha ) = q$. Given a live complete set $P$ for a target automaton $\Aut$, the 
essential idea of the $\ID$
algorithm is to first construct the set $T'=P \cup \{ f(\alpha,b)\vert(\alpha,b)\in P \times \Sigma \} \cup \{ d_0 \}$
of all one element extensions of strings in $P$ as a set of state names for the hypothesis automaton. 

A symbol $d_0$ is added as a name for the canonical dead state. Now this dead state can be used for incremental learning of a DFA, since parts of the DFA which have not yet been learned 
can be "hidden" inside the dead state. This is a key idea in the $\IID$ incremental learning algorithm described in 
Section 4.2

The 
set  of state names is then iteratively partitioned into sets $E_i(\alpha) \subseteq T'$ for $i = 0, 1, \ldots$ such that elements 
$\alpha$, $\beta$ of $T'$ that
denote the same state in $\Aut$ will occur in the same partition set, i.e. $E_i(\alpha) = E_i(\beta)$. 
This partition refinement can be proven to terminate and the resulting collection of 
sets forms a congruence on $T'$. Finally the $\ID$ algorithm constructs the hypothesis DFA as 
the resulting quotient DFA.
The method used to refine the partition set is to iteratively construct a set $V$ of {\it distinguishing strings},  such that no two distinct states of $\Aut$ have the same behaviour on all of $V$.

\begin{algorithm}
\singlespacing
\begin{raggedright}
\textbf{Input}: A live complete set $P \subseteq \Sigma^*$ and a DFA ${\Aut}$ to act as a teacher 
answering membership queries $\alpha\in L({\Aut})?$\\
\textbf{Output:} A $DFA$ $M$ language equivalent to the target DFA ${\Aut}$.
\par\end{raggedright}

\begin{enumerate}
\item \textbf{begin }
\item \textbf{/}/Perform Initialization
\item $i=0,$ $v_i=\lambda,$ $V=\lbrc v_i \rbrc,$ 
\item $T=P \cup \{f(\alpha,b)\vert(\alpha,b)\in P \times \Sigma\},$  
$T'=T \cup \lbrc d_0 \rbrc$
\item Construct function  $E_{0}$ for $v_0=\lambda,$
\item $E_{0}(d_{0})=\emptyset$
\item $\forall\alpha\in T$ 
\item \{ pose the membership query $``\alpha\in L({\Aut})?"$
\item \textbf{$\qquad$if} the teacher's response is $yes$
\item \textbf{$\qquad$then} $E_{0}(\alpha)=\{\lambda\}$
\item \textbf{$\qquad$else} $E_{0}(\alpha)=\emptyset$
\item \textbf{$\qquad$end if}
\item \}
\item //Refine the partition of the set $T'$ 
\item \textbf{while} $(\exists\alpha,\beta\in P^{\prime}$ and $b\in\Sigma$
such that\\ $E_{i}(\alpha)=E_{i}(\beta)$ but $E_{i}(f(\alpha,b))\not\neq E_{i}(f(\beta,b)))$ 
\item \textbf{do}
\item $\qquad$Let $\gamma\in E_{i}(f(\alpha,b))\oplus E_{i}(f(\beta,b))$
\item $\qquad$$v_{i+1}=b\gamma$
\item $\qquad$$V=V\cup\{v_{i+1}\}$, $i=i+1$
\item $\qquad$$\forall\alpha\in T_{k}$ pose the membership query $"\alpha v_{i}\in L({\Aut})?"$
\item $\qquad\qquad$\{
\item \textbf{$\qquad$$\qquad$if} the teacher's response is $yes$ 
\item \textbf{$\qquad$$\qquad$then} $E_{i}(\alpha)=E_{i-1}(\alpha)\cup\{v_{i}\}$
\item \textbf{$\qquad\qquad$else} $E_{i}(\alpha)=E_{i-1}(\alpha)$
\item \textbf{$\qquad\qquad$end if}
\item $\qquad\qquad$\}
\item \textbf{{end while}}
\item //Construct the representation $M$ of the target DFA ${\Aut}$.
\item The states of $M$ are the sets $E_{i}(\alpha)$, where $\alpha\in T$ 
\item The initial state $q_{0}$ is the set $E_{i}(\lambda)$
\item The accepting states are the sets $E_{i}(\alpha)$ where $\alpha\in T$
and $\lambda\in E_{i}(\alpha)$
\item The transitions of $M$ are defined as follows:
\item $\quad\forall\alpha\in P^{\prime}$
\item \textbf{$\qquad$if} $E_{i}(\alpha)=\emptyset$
\item \textbf{$\qquad$then} add self loops on the state $E_{i}(\alpha)$
for all $b\in\Sigma$
\item \textbf{$\qquad$else} $\forall b\in\Sigma$ set the transition $\delta(E_{i}(\alpha),b)=E_{i}(f(\alpha,b))$
\item \textbf{$\qquad$end if}
\item \textbf{{end.}}
\end{enumerate}
\caption{\label{IDalg} \textbf{$\ID$ Learning Algorithm}}
\end{algorithm}

In Section 4.2, we present the DFA family learning algorithm $\IID$ so that 
similar variables in the $\IID$ and $\ID$ algorithms share similar names.
This pedagogic device emphasises some similarity in the behaviour of both algorithms. However, there are also important 
differences of behaviour. Thus, when analysing the behavioural properties of similar program variables, we will try to distinguish their context as $v_n^\ID$, 
$\EID {n} \alpha, \ldots $ etc, (for the $\ID$ algorithm) and correspondingly $v_n^c$, 
$E_{n}^c ( \alpha ), \ldots$ etc, (for the $\IID$ algorithm).
Our basic argument in the proof of correctness of $\IID$ is to show how the learning behaviour of $\IID$ on a sequence of input strings $s_1 \c \ldots s_n \in \Sigma^*$ can be simulated by the behaviour of $\ID$ on the prefix closure 
$\Pref  ( \lbrc s_1 \c \ldots s_n \rbrc )$ of the 
corresponding set of inputs $\lbrc s_1 \c \ldots s_n \rbrc$. Once this is established one can apply the 
correctness of $\ID$ to establish the correctness of $\IID$.  The correctness of the
$\ID$ algorithm can be stated as follows. 

\startdef
\noindent {\bf 4.1.1. Theorem.}\\
{\sl
\noindent (i) Let $P \subseteq \Sigma^*$ be a live complete set for a DFA $\Aut$ containing $\lambda$. Then given $P$ and $\Aut$ as input, the $\ID$ algorithm 
terminates and the automaton $M$ returned  is the canonical minimum state automaton for $L(\Aut)$.\\
\noindent (ii) Let $l \in \Nat$ be the maximum value of program variable $i^{\ID}$ given $P$ and $\Aut$. 
For all $0 \leq n \leq l$ and for all $\alpha \in T$,
$$
\EID n \alpha = \lbrc v_j^{\ID} \in V^{\ID}  \st 0 \leq j \leq n , \sp \alpha v_j^{\ID}  \in L(\Aut) \rbrc .
$$
}
\endef
\noindent {\bf Proof.} 
\noindent (i) See  \cite{AngluinID} Theorem 3.\\
\noindent (ii) By induction on $n$. 

\noindent {\bf Basis.} Suppose $n = 0$. Then $v_0^{\ID} = \lambda$. For any $\alpha \in T$, 
if $\alpha v_0^{\ID}  \in L(\Aut)$ then $\alpha \in L(\Aut)$ so 
$\EID 0 \alpha = \lbrc  v_0^{\ID} \rbrc$. If 
$\alpha v_0^{ID} \not\in L(\Aut)$ then $\alpha  \not\in L(\Aut)$ so 
$\EID 0 \alpha = \emptyset$.
Thus 
$
\EID 0 \alpha = \lbrc v_j^{\ID} \st 0 \leq j \leq 0 , \sp \alpha v_j^{ID}  \in L(\Aut) \rbrc 
$.

\noindent {\bf Induction Step.} Suppose $l \geq n > 0$. Consider any $\alpha  \c \beta \in P'$ and 
$b \in \Sigma$ such that 
$\EID {n-1} \alpha = \EID {n-1} \beta$ but 
$\EID {n-1} { f(\alpha, b)} \not= \EID {n-1} { f(\beta, b)}$.
Since $n-1 < l$ then $\alpha$, $\beta$ and $b$ exist.
Then 
$$
\EID {n-1} { f(\alpha, b)} \disjointsum \EID {n-1} { f(\beta, b)} \not= \emptyset .
$$
Consider any $\gamma \in \EID {n-1} { f(\alpha, b)} \disjointsum \EID {n-1} { f(\beta, b)}$
and let $v_n^{\ID} = b \gamma$. 
For any $\alpha \in T$, if $\alpha v_n^{\ID}  \in L(\Aut)$ then 
$\EID {n} \alpha = \EID {n-1} \alpha \cup \lbrc v_n^{\ID} \rbrc$
and if $\alpha v_n^{\ID}  \not\in L(\Aut)$ then 
$\EID {n} \alpha = \EID {n-1} \alpha$. So by the induction hypothesis
$
\EID n \alpha = \lbrc v_j^{\ID} \in V^{\ID}  \st 0 \leq j \leq n , \sp \alpha v_j^{\ID}  \in L(\Aut) \rbrc 
$.
\endproof

\subsection{The $\IID$ Algorithm}

We can now present the $\IID$ algorithm for incremental lazy learning of a family of DFA. 
We  give a rigorous proof that  $\IID$ correctly learns in the limit 
in the sense of \cite{Gold} (Correctness Theorem 4.2.6). 

Algorithm \ref{KIIDalg} is the main component of the $\IID$ algorithm. It learns a sequence 
$F_0 \c \ldots \c F_l$ of families $F_i = ( M_i^1 \c \ldots \c M_i^n )$ of $n$  
DFA driven by a sequence 
of input strings (queries) $s_1 \c \ldots \c  s_l $. 
The teacher is a single $n$-bit Kripke structure ${\Aut}$.
Then $F_0$ is a null hypothesis about  the projections ${\Aut}_1  \c \ldots \c {\Aut}_n$ of ${\Aut}$.
We claim that the sequence $F_0 \c F_1  \c \ldots$ finitely converges to  the projections $( {\Aut}_1  \c \ldots \c {\Aut}_n )$ given enough information about ${\Aut}$, i.e when $s_1 \c \ldots \c  s_l $ contains a live complete set of queries for each 
projection ${\Aut}_i$.

\begin{algorithm}
\begin{raggedright}
\textbf{Input}: A file $S = s_1 \c \ldots \c  s_l $ of input strings $s_i \in \Sigma^*$ and an $n$-bit Kripke structure ${\Aut}$ 
as teacher to answer queries $\lambda_{\Aut}^* (  s_i ) = \smsp ?$\\
\textbf{Output:} A sequence of families  $F_{t} = ( M_t^1 \c \ldots \c M_t^n )$ of DFA for $t = 0 \c \ldots \c l$. 
\par\end{raggedright}
\begin{enumerate}
\item \textbf{begin }
\item $\quad$ \textbf{/}/Perform Initialization
\item $\quad$ \textbf{for} c = 1 to n \textbf{do} $\lbrc$ $i_c=0,$  $v_{i_c}^c = \varepsilon$, $V_c = \{  v_{i_c}^c \} \rbrc  $
\item $\quad$ $k=0,$ $t=0,$

\item $\quad$ $P_{0}=\{ \varepsilon  \}$, $P'_{0}=P_0 \cup \{ d_0  \}$, $T_{0}=P_0 \cup \Sigma$
\item $\quad$ //Build equivalence classes for the dead state $d_0$
\item $\quad$ \textbf{for} c = 1 to n \textbf{do} $\{ \sp E_{0}^c (d_{0})=\emptyset \sp \} $
\item $\quad$ //Build equivalence classes for input strings of length zero and one 
\item $\quad$ $\forall\alpha \in T_{0}$ $\{$
\item $\qquad$$(b_1  \c \ldots \c  b_n ) = \lambda_{\Aut}^* ( \alpha  )$
\item $\qquad$ \textbf{for} c = 1 to n \textbf{do} 
\item $\quad$ $\qquad$\textbf{if}  $b_c$ \textbf{then} $E_{i_c}^c(\alpha)=\{  v_{i_c}^c \}$ \textbf{else} $E_{i_c}^c(\alpha)=\emptyset$
\item $\quad$\}
\item $\quad$//Refine the initial equivalence relations $E_0^1 \c \ldots \c  E_0^n$  
\item $\quad$//into congruences using Algorithm \ref{Krefpartalg} 
\item $\quad$
\item $\quad$//Synthesize an initial family $F_{0}$ approximating ${\Aut}$
\item $\quad$//using Algorithm \ref{Kautconstalg}.
\item $\quad$
\item $\quad$//Process the file of examples.
\item $\quad$\textbf{while } $S \not= empty$ \textbf{do} $\{$
\item $\qquad$read( S, $\alpha$ )
\item $\qquad$k = k+1, t = t+1
\item $\qquad$$P_{k}=P_{k-1}\cup Pref(\alpha)$ //prefix closure 
\item $\qquad$$P_{k}^{\prime}=P_{k}\cup\{d_{0}\}$
\item $\qquad$$T_{k} =P_{k}  \cup\{ f( \alpha \c b ) \st \alpha \in P_{k} - P_{k-1}, b \in \Sigma\}$ //for prefix closure
\item $\qquad$$T'_{k} = T_{k} \cup \{ d_{0} \}$
\item $\qquad$$\forall\alpha\in T_{k} -  T_{k-1}  \sp \{ $ 
\item $\qquad$$\qquad$ \textbf{for} c = 1 to n \textbf{do} $E_{0}^c(\alpha) = \emptyset$ //initialise the new equivalence class $E_{0}^c(\alpha)$
\item $\qquad$$\qquad$ \textbf{for} j = 0 to $i_c$ \textbf{do} $ \{ $
\item $\qquad$$\qquad$$\qquad$// Consider adding previous distinguishing string $v_{j}^c \in V_c$ 
\item $\qquad$$\qquad$$\qquad$// to the new equivalence class $E_{j}^c(\alpha)$
\item $\qquad$$\qquad$$\qquad$$(b_1  \c \ldots \c  b_n ) = \lambda_{\Aut}^* ( \alpha  \concat v_{j}^c )$
\item $\qquad$$\qquad$$\qquad$\textbf{if}  $b_c$ \textbf{then} $E_{j}^c(\alpha) = E_{j}^c(\alpha) \cup \lbrc v_{j}^c \rbrc$
\item $\qquad$$\qquad$\}
\item $\qquad$\}
\item $\qquad$//Refine the current equivalence relations $E_{i_1}^1 \c \ldots \c  E_{i_n}^n$  
\item $\qquad$// into congruences using Algorithm \ref{Krefpartalg} 
\item $\qquad$
\item $\qquad$\textbf{if} $\alpha$ is consistent with $F_{t-1}$
\item $\qquad$\textbf{then} $F_{t}=F_{t-1}$ 
\item $\qquad$\textbf{else} synthesize the family $F_{t}$ using Algorithm \ref{Kautconstalg}.
\item $\qquad$$\}$
\item \textbf{end.}
\end{enumerate}
\caption{\label{KIIDalg} \textbf{FID: a DFA Family Learning Algorithm}}
\end{algorithm}
The basic idea of Algorithm \ref{KIIDalg} is to construct 
in parallel a
family 
$$( E_{i_1}^1 \c \ldots \c  E_{i_n}^n ) $$ 
of $n$ individual equivalence relations on the same set $T_k$ of state names.
For each equivalence relation $E_{i_j}^j$, a set $V_j$ of distinguishing strings is incrementally generated  
to split pairs of equivalence classes in $E_{i_j}^j$ until a congruence is achieved.
Then a quotient DFA $M^j$ can be constructed from the partition of $T_k$ by the congruence $E_{i_j}^j$.
The congruences are 
constructed so that  $E_i^j \subseteq E_{i+1}^j$ and thus the $\IID$ algorithm is incremental, and fully reuses 
information about previous approximations, which is efficient.

Each  DFA family $F_i = ( M_i^1 \c \ldots \c M_i^n )$ 
is constructed from the partition family $ ( E_{i_1}^1 \c \ldots \c  E_{i_n}^n )$
using Synthesis Algorithm \ref{Kautconstalg}.  When the $\IID$ algorithm is applied to the problem of LBT, 
the input strings $s_{i} \in \Sigma^*$ to $\IID$ are generated as counterexamples to correctness (i.e. test cases). 
For this we execute a model checker on a Kripke structure ${\Aut}_{i-1} $ which is a minimised subdirect product of 
$( M_{i-1}^1 \c \ldots \c M_{i-1}^n )$
using a requirements specification $\phi$ expressed in temporal logic. 
(The construction of  ${\Aut}_{i-1} $ will be detailed in Sections 5 and 6.)
In the case that no counterexamples to $\phi$ can be found 
in ${\Aut}_{i-1} $ then $s_{i}$ is randomly chosen, taking care to avoid all previously used input strings.

Algorithm \ref{Krefpartalg} implements lazy partition refinement, to extend 
$E_{i_1}^1 \c \ldots \c  E_{i_n}^n$
from being equivalence relations on states to being
a family of congruences with respect to the state transition functions $\delta_{1}  \c \ldots \c \delta_{n}$
for the synthesized DFA $M^1\c \ldots \c M^n$.

Thus line 1 in Algorithm \ref{Krefpartalg} searches for congruence failure in any one of the equivalence relations 
$E_{i_1}^1 \c \ldots \c  E_{i_n}^n$.
In lines 6-14 of Algorithm \ref{Krefpartalg} we apply lazy partition refinement. This technique implies reusing the new distinguishing string $v$ wherever possible to refine each equivalence relation $E_{i_j}^j$ that is not yet a congruence. On the other hand, any 
equivalence relation $E_{i_j}^j$ that is already a congruence is not refined, even though the result $b_j$ of the 
new query 
$\alpha  \concat v$ might add some new information to $M^j$. This brings the set of relations 
$E_{i_1}^1 \c \ldots \c  E_{i_n}^n$ to a simultaneous fixed point of $n$ congruence constructions as soon as possible.
It therefore helps to reduce the number of active learner queries and raise the number of model checker queries used during learning based testing (cf. Section 1.2).

\begin{algorithm}
\begin{enumerate}
\item \textbf{while} $(\exists \smsp 1 \leq c \leq n, \exists\alpha,\beta\in P_{k}^{\prime}$ and $\exists \sigma \in \Sigma$
such that $E_{i_c}^c(\alpha)=E_{i_c}^c(\beta)$ but $E_{i_c}^c( f( \alpha \c \sigma ) )\not\neq E_{i_c}^c( f( \beta \c \sigma))$ \textbf{do} $ \{ $
\item $\quad$//Equivalence relation $E_{i_c}^c$ is not a congruence w.r.t. $\delta_c$
\item $\quad$//so add a new distinguishing sequence.
\item $\quad$\textbf{Choose} $\gamma\in E_{i_c}^c( f( \alpha \c \sigma ) )\oplus E_{i_c}^c( f( \beta \c \sigma ) )$
\item $\quad$$v_{}=\sigma \concat \gamma$ 
\item $\quad$$\forall\alpha\in T_{k}$ $\{$
\item $\qquad$$(b_1  \c \ldots \c  b_n ) = \lambda_A^* ( \alpha  \concat v_{})$
\item $\qquad$\textbf{for} c = 1 to n \textbf{do} $ \{ $
\item $\quad$\textbf{$\qquad$if} $E_{i_c}^c(\alpha)=E_{i_c}^c(\beta)$ and $E_{i_c}^c( f( \alpha \c \sigma ) )\not\neq E_{i_c}^c( f( \beta \c \sigma ))$ \textbf{then} $\{$
\item $\qquad$$\qquad$// Lazy refinement of equivalence relation $E_{i_c}^c$
\item $\qquad$$\qquad$$i_c=i_c+1$, $v_{i_c} = v$, $V_c=V_c\cup\{v_{i_c}\}$
\item $\qquad$$\qquad$\textbf{if}  $b_c$ \textbf{then} $E_{i_c}^c(\alpha)=E_{i_c - 1}^c(\alpha)\cup\{  v_{i_c} \}$ \textbf{else} $E_{i_c}^c(\alpha)=E_{i_c - 1}^c(\alpha)$
\item $\qquad$$\quad$$\}$
\item $\qquad$\}
\item $\}$
\end{enumerate}
\caption{\label{Krefpartalg} Lazy Partition Refinement}
\end{algorithm}

\begin{algorithm}
\begin{enumerate}
\item \textbf{for} c = 1 to n \textbf{do} $\{ \sp$
\item $\qquad$// Synthesize the quotient DFA  $M^c$
\item $\qquad$ The states of $M^c$ are the sets $E_{i_c}^c(\alpha)$, where $\alpha\in T_{k}$ 
\item $\qquad$ Let $q_{0}^{c} = E_{i_c}^c(\varepsilon)$
\item $\qquad$ The accepting states are the sets $E_{i_c}^c(\alpha)$ where $\alpha\in T_{k}$
and $\varepsilon \in E_{i_c}^c(\alpha)$
\item $\qquad$ The transition function $\delta_c$ of $M^c$ is defined as follows:
\item $\qquad$$\quad$$\forall\alpha\in P_{k}^{\prime}$ $\{$
\item $\qquad$ $\qquad$ \textbf{if} $E_{i_c}^c(\alpha)=\emptyset$ \textbf{then} $\forall b\in\Sigma$ $\{$ let $\delta_c (E_{i_c}^c(\alpha),b)=E_{i_c}^c(\alpha)$ $\}$
\item $\qquad$ $\qquad$ \textbf{else} $\forall b\in\Sigma$ $\{$ $\delta_c (E_{i_c}^c(\alpha),b)=E_{i_c}^c(\alpha \concat b)$ $\}$
\item $\qquad$$\quad$$\}$
\item $\qquad$  $\quad$$\forall \beta\in T_{k} - P_{k}^{\prime}$ $\{$
\item $\qquad$$\qquad$$\quad$\textbf{if} $\forall\alpha\in P_{k}^{\prime}$
$\{$ $E_{i_c}^c(\beta)\neq E_{i_c}^c(\alpha)$ $\}$ and $E_{i_c}^c(\beta)\neq\emptyset$ \textbf{then} 
\item $\qquad$$\qquad$$\qquad$$\forall b\in\Sigma$ $\{$ $\delta_c ( E_{i_c}^c(\beta), b) = \emptyset$ $\}$
\item $\qquad$$\quad$$\}$
\item $\}$
\item \textbf{return} $F = ( M^1 \c \ldots \c M^n )$
\end{enumerate}
\caption{\label{Kautconstalg} DFA Family Synthesis}
\end{algorithm}



We begin an analysis of the correctness of the $\IID$ algorithm by confirming that the 
construction of hypothesis DFA carried out by Algorithm \ref{Kautconstalg}
is well defined. 

\startdef
\noindent {\bf 4.2.1. Proposition.}
{\sl
For each $t \geq 0$ the hypothesis DFA $M^1 \c \ldots \c M^n$ constructed by the DFA Family Synthesis Algorithm \ref{Kautconstalg} after $t$ input strings have been applied to ${\Aut}$ are all well defined DFA. 
}
\endef
\noindent {\bf Proof.} 
The main task is to show $\delta$  to be well defined function and uniquely
defined for every state $E_{i}(\alpha)$, where $\alpha\in T_{k}$.
\endproof

Proposition 4.2.1 establishes that Algorithm \ref{Kautconstalg} will generate families of well defined DFA. However, to show that the
$\IID$ algorithm learns correctly in the limit, we must prove that this sequence of DFA families finitely converges to the $n$ individual projections 
$\Aut_i$ of the target Kripke structure $\Aut$. It will suffice to show that the behaviour of $\IID$ can be simulated by the behaviour of $\ID$, since $\ID$ is known to learn correctly given a live complete set of input strings (c.f. Theorem 4.4.1.(i)). The first step in this proof is to show that the sequences of sets of state names $\PIID k$ and $\TIID k$ generated by $\IID$ converge to the sets $\PID$ and $\TID$ of $\ID$.

\startdef
\noindent {\bf 4.2.2. Proposition.}
{\sl
Let $S = s_1 \c \ldots \c s_l$ be any non-empty sequence of input strings $s_i \in \Sigma^*$ for $\IID$
and let $\PID = \Pref ( \lbrc \lambda \c s_1 \c \ldots \c s_l \rbrc ) $ be the prefix closure of the corresponding input set for $\ID$.
\vskip 5pt

\noindent (i) For all $0 \leq k \leq l$, 
$\PIID k = \Pref ( \lbrc \lambda \c s_1 \c \ldots \c s_k \rbrc ) \subseteq \PID$.

\noindent (ii) For all $0 \leq k \leq l$, 
$\TIID k = \PIID k \cup \lbrc f( \alpha \c b ) \st \alpha \in \PIID k \c b \in \Sigma \rbrc  \subseteq \TID$.

\noindent (iii) $\PIID l = \PID$ and $\TIID l = \TID$.

\label{lockstep}
}
\endef
\noindent {\bf Proof.} 
Clearly (iii) follows from (i) and (ii). Then (i) and (ii) are easily proved by induction on $k$. 
\endproof
Observe that unlike $\IID$, the $\ID$ algorithm does not 
compute any prefix closure of input strings. Therefore, prefix closure must be added explicitly in 
Proposition 4.2.2, to make a correspondence between the behaviour of $\IID$ and $\ID$.

Next we turn our attention to proving some fundamental loop invariants for Algorithm  \ref{KIIDalg}. Since this algorithm in turn calls the Lazy Partition Refinement Algorithm \ref{Krefpartalg} then we have in effect a doubly nested loop structure to analyse. Clearly the 
outer loop counter $k$ in Algorithm  \ref{KIIDalg} and the family of inner loop counters $i_c$ (for $1 \leq c \leq n$) in 
Algorithm \ref{Krefpartalg} both increase on each iteration. 
However, the relationships between these counter variables are not easily defined. 
Nevertheless, since all variables increase from an initial value of zero, we can assume the existence of some family 
of $n$ monotone re-indexing functions that capture their relationships.

\startdef
\noindent {\bf 4.2.3. Definition.}
Let $S = s_1 \c \ldots \c s_l$ be any non-empty sequence of strings $s_i \in \Sigma^*$.
The {\it re-indexing function}
$\Kfunc_c^S : \Nat \to \Nat$ for  $\IID$ on input $S$ (for each $1 \leq c \leq n$) is the unique monotonically increasing function 
such that for each $n \in \Nat$, $\Kfunc_c^S (n)$ is the least 
integer $m$ such that program variable $k$ has value $m$ while the program variable $i_c$ has value $n$. 
Thus, for example, $\Kfunc_c^S (0) = 0$ for all $1 \leq c \leq n$. When $S$ is clear from the context, we may simply write $\Kfunc_c$ for $\Kfunc_c^S$.
\endef

With the help of these re-indexing functions we can express important invariant properties of the 
distinguishing sequence variables 
$v_j^c$ and partition set variables $E_n^c ( \alpha)$. Using Proposition 4.2.2 their relationship to 
the corresponding variables $v_j^\ID$ and $\EID n \alpha$ of $\ID$ can be established.
Since Algorithm  \ref{KIIDalg} has a doubly nested loop structure, the proof of Simulation Theorem 4.2.4 below makes  use of a doubly nested induction argument. 

\startdef
\noindent {\bf 4.2.4. Simulation Theorem.}
{\sl
Let $S = s_1 \c \ldots \c s_l$ be any non-empty sequence of strings $s_i \in \Sigma^*$.
For any execution of  $\IID$ on $S$ and the $n$-bit Kripke structure ${\Aut}$ there exists an execution of $\ID$ on 
$\Pref ( \lbrc \lambda \c s_1 \c \ldots \c s_l \rbrc )$ and the $c$-th projection ${\Aut}_c$ (for each $1 \leq c \leq n$) such that for all 
$m \geq 0$:

\noindent (i) For all $n \geq 0$, if $\Kfunc_c(n) = m$ then:

\indent \hskip 14pt (a) for all $0 \leq j \leq n$, $v_j^c = v_j^\ID$,

\indent  \hskip 14pt (b) for all $0 \leq j < n$, $v_n^c \not= v_j^c$,

\indent \hskip 14pt (c) for all $\alpha \in \TIID m$,
$E_n^c ( \alpha ) = \lbrc v_j^c \in V_c \st 0 \leq j \leq n , \sp \alpha v_j^c  \in L({\Aut}_c) \rbrc$.

\noindent (ii) If $m > 0$ then let $p \in \Nat$ be the greatest integer such that $\Kfunc_c(p) = m-1$. 
Then for all $\alpha \in \TIID m$,
$E_{p}^c ( \alpha ) = \lbrc v_j^c \in V_c \st 0 \leq j \leq p , \sp \alpha v_j^c  \in L({\Aut}_c) \rbrc$.

\noindent (iii) The $m$th partition refinement of $\IID$ terminates.
}
\endef
\noindent {\bf Proof.} By induction on $m$ using Proposition 4.2.2.(i).
\endproof

Part (i.a) above asserts that the same distinguishing sequences are produced in the same order by $\IID$ and $\ID$. 
Part (i.b) asserts that a distinguishing sequence is never produced twice by $\IID$.
Part (i.c) and (ii) characterise the partition sets $E_n^c ( \alpha )$ as sets of  all distinguishing sequences $v_j^c$ that lead to an accepting state of ${\Aut}_c$ from $\alpha$.

Note that both $\ID$ and $\IID$ are non-deterministic algorithms (due to the non-deterministic choice on line 17 of Algorithm \ref{IDalg} and line 4 of Algorithm \ref{Krefpartalg}).
Therefore in the statement of Theorem 4.2.4 above,  we can only talk about the existence of {\it some} correct simulation. Clearly there are also simulations of $\IID$ by $\ID$ which are not correct, but this does not affect the basic correctness argument.  

\startdef
\noindent {\bf 4.2.5. Corollary.}
{\sl
Let $S = s_1 \c \ldots \c s_l$ be any non-empty sequence of strings $s_i \in \Sigma^*$.
Any execution of $\IID$ on $S$ and an $n$-bit Kripke structure ${\Aut}$ terminates with the program variable $k$ having value $l$.
}
\endef
\noindent {\bf Proof.} Follows from Simulation Theorem 4.2.4.(iii) since clearly the while loop of  
Algorithm  \ref{KIIDalg} terminates when the input sequence $S$ is empty.
\endproof

Using the detailed analysis of the invariant properties of the 
program variables $\PIID k$ and $\TIID k$ in Proposition 4.2.2 and 
$v_j^c$ and $E_n^c (\alpha)$ in Simulation Theorem 4.2.4 it is now a simple matter to establish correctness of learning
for the $\IID$ Algorithm.

\startdef
\noindent {\bf 4.2.6. Correctness Theorem.}
{\sl
Let $S = s_1 \c \ldots \c s_l$ be any non-empty sequence of strings $s_i \in \Sigma^*$
such that $\lbrc  \lambda \c s_1 \c \ldots \c s_l \rbrc$ contains a live complete set for each projection $\Aut_i$ of $\Aut$.
Then $\IID$ terminates on $S$. Also for each $1 \leq i \leq n$ the hypothesis
DFA $M_{l}^i$ is a canonical representation of $\Aut_i$ .
}
\endef
\noindent {\bf Proof.} By Corollary 4.2.5, $\IID$ terminates on $S$ with the variable $k$ having value $l$.
By Simulation Theorem 4.2.4.(i) and Theorem 4.1.1.(ii), there exists an execution of $\ID$ on 
$\Pref ( \lbrc \lambda \c s_1 \c \ldots \c s_l \rbrc )$ 
such that
$E_{n}^i ( \alpha ) = \EID {n} \alpha$ for all $\alpha \in \TIID {l}$ and any $n$ such that 
$\Kfunc (n) = l$. 
By Proposition 4.2.2.(iii), $\TIID {l} = \TID$ and $\PIIDprime {l} = \PIDprime$. So letting $M_i^\ID$ be the canonical 
representation of $\Aut_i$
constructed by $\ID$ using  $\Pref ( \lbrc \lambda \c s_1 \c \ldots \c s_l \rbrc )$
then $M_i^\ID$ and $M_{l}^i$ have the same state sets,  initial
states, accepting states and transitions. 
\endproof

Our next result  confirms that each hypothesis DFA $M_{t}^i$ generated after $t$ input strings
have been applied to ${\Aut}$ is consistent with all currently known observations about the $i$th projection $\Aut_i$.
This is quite straightforward in the light of Simulation Theorem 4.2.4. 

\startdef
\noindent {\bf 4.2.7. Compatibility Theorem.}
{\sl
Let $S = s_1 \c \ldots \c s_l$ be any non-empty sequence of strings $s_i \in \Sigma^*$. 
For each $0 \leq t \leq l$ 
and each string $s \in \lbrc  \lambda \c s_1 \c \ldots \c s_t \rbrc$, the hypothesis automaton 
$M_{t}^i$ accepts $s$ if, and only if  
the $i$th projection $\Aut_i$ of $\Aut$ does.
}
\endef
\noindent {\bf Proof.}  By definition, $M_{t}^i$ is compatible with $\Aut_i$ on 
$\lbrc  \lambda \c s_1 \c \ldots \c s_t \rbrc$ if, and only if, 
for each $0 \leq j \leq t$, 
$s_j \in L(\Aut_i) \sp \Leftrightarrow \sp \lambda \in E_ {i_t}^i  ({s_j})$,
where ${i_t}$ is the greatest integer such that $\Kfunc({i_t}) = t$ and the sets 
$ E_{i_t}^i  ({\alpha})$ for $\alpha \in \TIID t$ are the states of $M_{t}^i$.
Now $v_0 = \lambda$. So by Simulation Theorem 4.2.4.(i).(c), if $s_j \in L(\Aut_i)$ 
then $s_j v_0 \in L(\Aut_i)$ so $v_0 \in  \EIID {i_t}  {s_j}$, i.e. $\lambda \in  E_{i_t}^i  ({s_j})$,
and if
$s_j \not\in L(\Aut_i)$ 
then $s_j v_0 \not\in L(\Aut_i)$ so $v_0 \not\in  E_{i_t}^i  ({s_j})$, i.e. $\lambda \not\in  E_{i_t}^i  ({s_j})$.
\endproof

We have now established a reliable method for decomposing the problem of learning a $k$-bit Kripke structure $\Aut$
into the problem of learning a family of $k$ individual DFA. This approach supports projection, as defined in Section 
1.2 and Definition 2.2.

\section{Subdirect Product Construction}
We next turn our attention to problem 
of efficiently recombining a family of $k$ individual DFA (the projections) into a single $k$-bit deterministic Kripke structure. 
For this we use
a well known algebraic construction known as a subdirect product. Informally, a subdirect product of a family 
$F =  \lbrk A_i \st i \in I  \rbrk $ of algebraic structures, is any subalgebra of the direct product 
$\Pi F = \Pi_{i \in I} A_i$ which projects onto (surjectively) each of its co-ordinate algebras $A_i$. 
The subdirect product construction was introduced in \cite{Birkhoff44} as a universal decomposition method 
applicable to any algebraic structures.
The reader may consult \cite{MeinkeTucker} for basic facts about subdirect products and their universal properties.
A specific definition for deterministic Kripke structures is given below.

To begin with, we observe that for black-box testing it suffices to learn a Kripke structure up to behavioural equivalence.

\startdef
\noindent {\bf 5.1. Definition.}
{ Let ${\Aut}$ and ${\BAut}$ be $k$-bit Kripke structures over a finite input alphabet $\Sigma$. We say that ${\Aut}$ and 
${\BAut}$ are {\it behaviourally equivalent}, and write ${\Aut} \equiv {\BAut}$ if, and only if, for every finite input sequence
$ \sigma_1  \c \ldots \c  \sigma_{i} \in \Sigma^*$ we have 
$$
\lambda_{\Aut}^* (\smsp  \sigma_1  \c \ldots \c  \sigma_{i}  \smsp ) 
= 
\lambda_{\BAut}^* ( \smsp  \sigma_1  \c \ldots \c  \sigma_{i}  \smsp ) .
$$
}
Clearly, by the isomorphism identified in Section 2 between $1$-bit Kripke structures and DFA, for such structures 
we have ${\Aut} \equiv {\BAut}$ if, and only if, $L({\Aut}') = L({\BAut}')$. Furthermore, if $Min({\Aut})$ is the minimal subalgebra of 
${\Aut}$  then $Min({\Aut}) \equiv {\Aut}$.

A family of $k$ individual $1$-bit Kripke structures (DFA) can be combined into a single $k$-bit Kripke structure using the following instance of the subdirect product construction. 

\startdef
\noindent {\bf 5.2. Definition.}
{
Let ${\Aut}_1 \c \ldots \c {\Aut}_k$ be a family of 1-bit Kripke structures, 
$$
{\Aut}_i = ( \smsp Q_i  \c \Sigma  \c \delta_i : Q_i  \mult  \Sigma  \to Q_i  \c q_i^0 \c  \lambda_i : Q \to \Bool \smsp ) 
$$ 
for $i = 1 \c \ldots \c  k$. 
Define the 
{\it direct product Kripke structure}
$$
\prod_{i = 1}^{k} {\Aut}_i = 
( \smsp Q  \c \Sigma  \c \delta : Q  \mult  \Sigma  \to Q  \c q^0 \c  \lambda : Q \to \Bool^k \smsp ),
$$
where
$Q = \prod_{i = 1}^{k} Q_i = Q_1 \mult  \ldots \mult Q_k $ and
$q^0 =  ( \smsp q_1^0 \c \ldots \c  q_k^0  \smsp )$. Also
$$
\delta ( q_1 \c \ldots \c  q_k \c  \sigma ) = 
( \smsp \delta_1 ( q_1 \c  \sigma ) \c \ldots \c \delta_k (  q_k \c  \sigma ) \smsp ) ,
$$
$$
\lambda ( q_1 \c \ldots \c  q_k ) = (  \smsp \lambda_1 ( q_1 ) \c \ldots \c \lambda_k (  q_k  ) \smsp ) .
$$
Associated with the direct product $\prod_{i = 1}^{k} {\Aut}_i$ we have $i$-th {\it projection mapping}
$$
\proj_i : Q_1 \mult  \ldots \mult Q_k \to Q_i
\c \quad
\proj_i ( q_1 \c \ldots \c  q_k ) = q_i
\c 
1 \leq i  \leq k
$$

Define the {\it subdirect product} $\Min ( \smsp \prod_{i = 1}^{k} {\Aut}_i \smsp )$ be the minimal subalgebra of 
$\prod_{i = 1}^{k} {\Aut}_i$.
}
\endef
The reason for taking the subdirect product of the ${\Aut}_i$ as the 
minimal subalgebra of the direct product $\prod_{i = 1}^{k} {\Aut}_i$ is to avoid the state space explosion due to a large number of unreachable states in the direct product itself. The state space size of $\prod_{i = 1}^{k} {\Aut}_i$ grows exponentially with $k$. On the other hand, since most of these
states are unreachable from the initial state, then from the point of view of requirements testing they are irrelevant. This subdirect product can be computed from its components ${\Aut}_i$ in time  $O( k. m . \vert  \Sigma \vert )$
where $m$ is the number of states in the resulting subdirect product and  $\vert  \Sigma \vert$ is the size of the input alphabet. 
A naive algorithm based on systematic path exploration starting from the initial state can be used. We leave the definition 
of this algorithm as an exercise for the reader.

As is well known from universal algebra, the $i$-th projection mapping $\proj_i$ is a homomorphism.
\startdef
\noindent {\bf 5.3. Proposition.}
{\sl Let ${\Aut}_1 \c \ldots \c {\Aut}_k$ be any minimal 1-bit Kripke structures. \\
\noindent {\bf (i)} For each $1 \leq i  \leq k$, the projection mapping $\proj_i : \Min ( \smsp \prod_{i = 1}^{k} {\Aut}_i \smsp ) \to {\Aut}_i$
is an epimorphism. Hence $\Min ( \smsp \prod_{i = 1}^{k} {\Aut}_i \smsp )$ is a subdirect product of the ${\Aut}_i$.
\vskip 6pt
\noindent {\bf (ii)}  
$\Min ( \smsp \prod_{i = 1}^{k} {\Aut}_i \smsp ) \equiv \prod_{i = 1}^{k} {\Aut}_i .$
}
\startproof
\noindent {\bf Proof.} (i) Immediate since the ${\Aut}_i$ are minimal. (ii) Follows from the fact that $Min({\Aut}) \equiv {\Aut}$.
\endproof
The following theorem justifies bit-sliced learning of $k$-bit Kripke structures using conventional regular inference methods for 
a family of DFA. It constitutes the correctness argument for the subdirect product component of the IKL architecture, as presented 
in Section 3.
\startdef
\noindent {\bf 5.4. Theorem.}
{\sl 
Let ${\Aut}$ be a $k$-bit Kripke structure over a finite input alphabet $\Sigma$. Let 
${\Aut}_1 \c \ldots \c {\Aut}_k$ be the $k$ individual 1-bit projections of ${\Aut}$. For any 1-bit Kripke structures 
${\BAut}_1 \c \ldots \c {\BAut}_k$, if, ${\Aut}_1 \equiv {\BAut}_1   \smsp \& \ldots \&  \smsp {\Aut}_k \equiv {\BAut}_k$
then 
$$
{\Aut} \equiv \Min ( \smsp \prod_{i = 1}^{k} {\BAut}_i \smsp ).
$$
}
\noindent {\bf Proof.} Use Proposition 5.3.
\endproof
By Correctness Theorem 4.2.6, the assumptions of Theorem 5.4 on the 1-bit Kripke structures 
${\BAut}_1 \c \ldots \c {\BAut}_k$ are fulfilled by the IKL architecture, since these are the canonical representations of ${\Aut}_1 \c \ldots \c {\Aut}_k$. 
So by Theorem 5.4, the output of the IKL algorithm, after DFA family learning has converged and the subdirect product construction has been applied 
is a $k$-bit Kripke structure ${\BAut}$ that is behaviourally equivalent with the input Kripke structure ${\Aut}$. 

Despite the canonical DFA 
${\BAut}_1 \c \ldots \c {\BAut}_k$ being minimal, the reduced product ${\BAut}$ may still be much 
larger in state space size than ${\Aut}$. 
This can slow down the process of model checking the output of IKL considerably. So it is important to reduce the state space size of 
${\BAut}$ even further. This last step of the IKL algorithm will be discussed in the next section.

\section{Kripke Structure Minimisation.}
In this section we introduce an efficient algorithm for the minimisation of deterministic Kripke structures with 
$\mathcal O( \vert  \Sigma \vert . n  \log_{2} n)$ time complexity. Here $n$ is the state space size of the Kripke structure ${\Aut}$ and 
$\vert  \Sigma \vert$ is the 
size of its input alphabet. This algorithm is applied on the back end of the IKL learning algorithm 
in order to speed up model checking of the learned hypothesis automata during testing. 

To define a minimisation algorithm, we need to generalise the concepts of 
{\it right language} and {\it Nerode congruence} from DFA to deterministic Kripke structures. We then show how Hopcroft's 
DFA minimisation algorithm of  \cite{Hop71} can be generalised to compute the Nerode congruence $\equiv$
of a deterministic Kripke structure ${\Aut}$. 
The quotient Kripke structure ${\Aut} / \equiv$ is minimal and language equivalent to ${\Aut}$.
This fact is the final result needed to prove the correctness of the IKL architecture.
We will prove the correctness and complexity properties of our minimisation algorithm from 
first principles. 

\subsection{Minimal Deterministic Kripke Structures}

Let us consider a DFA ${\Aut} = ( Q, \Sigma, \delta, q_{0}, F )$ . For each state $q\in Q$ of ${\Aut}$ there corresponds a
subautomaton of ${\Aut}$ rooted at $q$ which accepts the regular language 
$\mathcal{L}_{q}({\Aut}) \subseteq \Sigma^*$, consisting of just those words accepted by the subautomaton with $q$ as initial state. 
Thus $\mathcal{L}_{q_0}({\Aut})$ is the language accepted by ${\Aut}$.
The language $\mathcal{L}_{q}({\Aut})$ is called either the \emph{future }of state \emph{q }or the \emph{right language }of \emph{q. }${\Aut}$ is \emph{minimal} (i.e. \emph{state minimal} as opposed to \emph{algebraically minimal}) if for each pair of distinct states \emph{$p,q\in Q$}, we have, $\mathcal{L}_{p}({\Aut})\neq\mathcal{L}_{q}({\Aut})$. For any regular language 
$\mathcal{L} \subseteq \Sigma^*$ there is a smallest DFA (in terms of the number of states) accepting $\mathcal{L}$. This DFA is minimal, and is 
unique up to isomorphism.

An equivalence relation $\equiv$ can be defined on 
the states of a DFA by $p\equiv q$ if and only if $\mathcal{L}_{p}({\Aut})=\mathcal{L}_{q}({\Aut})$.
This relation is a congruence, i.e. if $p \equiv q$  then 
$p. \sigma \equiv q. \sigma $ for all $\sigma \in\Sigma^{*}$. It is known as the \emph{Nerode congruence}. 
Consider the quotient DFA ${\Aut} / \equiv$. This is the unique smallest DFA 
which accepts the 
regular language $\mathcal{L}_{q_0}({\Aut})$.
The problem of minimizing a DFA ${\Aut}$ is therefore to compute its Nerode congruence, which will be the identity relation if, and only if ${\Aut}$ is a minimal automaton.

The problem of computing a minimal Kripke structure ${\Aut}$  is 
an analogous but more general problem. 
In this case, the right language $\mathcal{L}_{q}({\Aut})$ associated with a state $q$ of ${\Aut}$ can be defined by
$$
\mathcal{L}_{q}({\Aut}) = 
\lbrc (\sigma_{1},...,\sigma_{n} , a ) \in \Sigma^* \times  \mathbb{B}^{k} 
\smsp \vert \smsp \lambda_{q}^{*}(\sigma_{1},...,\sigma_{n}) = a \rbrc .
$$
As before, ${\Aut}$ is\emph{ minimal} if for each pair of distinct states \emph{$p,q\in Q$} we have, $\mathcal{L}_{p}({\Aut})\neq\mathcal{L}_{q}({\Aut})$. 
There is again a smallest Kripke structure  associated with a right language 
$\mathcal{L} \subseteq \Sigma^* \times  \mathbb{B}^{k}$. This Kripke structure is also minimal, and 
unique up to isomorphism.
The Nerode congruence for a Kripke structure ${\Aut}$ is now defined by:
\begin{center}
$p\equiv q$ if and only if  $\lambda_{p}^{*}(\sigma_{1},...,\sigma_{n})=\lambda_{q}^{*}(\sigma_{1},...,\sigma_{n})$ for all $(\sigma_{1},...,\sigma_{n})\in\Sigma^{*}$.
\par\end{center}
and 
${\Aut} / \equiv$ is the unique smallest Kripke structure 
associated with the 
right language $\mathcal{L}_{q_0}({\Aut})$.
So the problem of minimising ${\Aut}$ is to compute this congruence.

\subsection{A Kripke Structure Minimisation
Algorithm}

\begin{algorithm}

 \caption{\label{KMN} Kripke Structure Minimisation}
\KwIn{A deterministic Kripke structure ${\Aut}$ with no unreachable states and $k$ output bits.}
\KwOut{The Nerode congruence $\equiv$ for ${\Aut}$, i.e. equivalence classes of states for the 
minimized structure ${\Aut}_{min}$ behaviourally equivalent to ${\Aut}$.}
     

Create\nllabel{partition} an initial state partition
$P=\{B_{q}=\{q^{\prime}\in
Q\;\vert\;\lambda(q)=\lambda(q^{\prime})\}\;\vert\;
q\in Q\}$. Let $n=\vert P\vert$. Let $B_{1},...,B_{n}$ be an
enumeration of P.

\textbf{if} {$n=\vert Q\vert$} \textbf{then}  \textbf{go to} line \ref{termination}.

    \ForEach{$\sigma\in\Sigma$\nllabel{subpartition}}{
\For{$i\leftarrow 1$ \KwTo $n$ } {$B(\sigma, i)=\{q\in B_{i}\;\vert\; \exists r \in Q \;s.t\; 
\delta(r,\sigma) = q\}.$
 /*This constitutes the subset of states in block $B_i$ which have predecessors
through input $\sigma$. */}
   
    }
$count=n+1$;

 \ForEach{$\sigma\in\Sigma$}{\nllabel{initwait}
choose\nllabel{initwait1} all the subsets $B(\sigma, i)$ (excluding any empty subsets) 
and put their block numbers $i$ on a waiting list (i.e. an
unordered
set) $W(\sigma)$ to be processed.
}
\BlankLine
Boolean splittable = true;

\While{splittable}{\nllabel{while}
 \ForEach{$\sigma\in\Sigma$ }{

 \ForEach{$i\in$ $W(\sigma)$}{

 Delete i from $W(\sigma)$\nllabel{bodystart}

\For{$j\leftarrow 1$ \KwTo $count-1$ s.t. $\exists t \in B_{j}$
with $\delta(t,\sigma)\in B(\sigma,i)$}{
Create $B_{j}^{\prime}=\{t\in B_{j}\;\vert\;\delta(t,\sigma)\in
B(\sigma,i)\}$\nllabel{newBj}

\If{$B_{j}^{\prime}\subset B_{j}$}{
$ B_{count}=B_{j}-B_{j}^{\prime}$; \nllabel{block}   $ B_{j}=B_{j}^{\prime}$

\ForEach{$\sigma\in\Sigma$}{
$B(\sigma,count)=\{q\in  B(\sigma,j) \;\vert\;    q \in B_{count}  \}$;\nllabel{newacount}

$ B(\sigma,j)=\{q\in B(\sigma,j)
  \;\vert 
\;  q \in B_{j}
\}$ \nllabel{newaj}

\If{$ j\notin W(\sigma)$ and\nllabel{if} $0<\vert B(\sigma,j)\vert\le\vert B(\sigma,count)\vert$}{$ W(\sigma)=W(\sigma)\cup\{j\}$}
\Else{ $W(\sigma)=W(\sigma)\cup\{count\}$\nllabel{endif}}

}

$ count=count+1$;
}

}
}
}

splittable = false;

\ForEach{$\sigma\in\Sigma$}{
\If {$W(\sigma) \not= \emptyset$}{splittable=true; \nllabel{bodyend}}
}
}
Return \nllabel{termination}partition blocks
$B_{1},...,B_{count}$.
 \end{algorithm}
 
Algorithm \ref{KMN} presents an efficient algorithm to compute the Nerode congruence $\equiv$ of a 
deterministic Kripke structure ${\Aut}$, which is the same as the state set of the associated quotient 
Kripke structure ${\Aut} / \equiv$.
We will give a rigorous but simple proof of the correctness of this algorithm.
By means of a new induction argument, we have simplified
the correctness argument compared with \cite{Ber11} and \cite{Hop71}.
First let us establish termination of the algorithm by using an appropriate well-founded ordering for the main loop variant.

\startdef
\noindent {\bf 6.2.1. Definition.}
{\sl
\label{def:definition3}Consider any pair of finite sets of
finite sets $A=\{A_{1},...,A_{m}\}$
and $B=\{B_{1},...,B_{n}\}$. We define an ordering relation
$\leq$ on $A$ and $B$
by $A\leq B$ iff $\forall1\leq i\leq m$, $\exists1\leq j\leq n$
such that $A_{i}\subseteq B_{j}$. Define $A<B\iff A\leq B\;\&\;
A\neq B$.
Clearly $\leq$ is a reflexive, transitive relation. Furthermore $\leq$ is well-founded, i.e. 
there are no infinite
descending chains $A_{1}>A_{2}>A_{3}...$ , since $\emptyset$ is
the smallest element under $\leq$.
}

\startdef
\noindent {\bf 6.2.2. Proposition.}
{\sl
Algorithm \ref{KMN} always terminates.
}

\noindent {\bf Proof.} 
We have two cases for the termination of the algorithm as a result of the partition formed on line \ref{partition} of the algorithm: (1) when $n=\vert Q\vert$, and (2) when $n<\vert Q\vert$.

 Consider the case when $n=\vert Q\vert$ then each block in the partition corresponds to a state of the
given Kripke structure with a unique bit-label and hence in this case the algorithm will terminate on line \ref{termination} by providing the description of these blocks.

Now consider the case when $n<\vert Q\vert$. Then the waiting sets $W(\sigma)$ for all $\sigma\in\Sigma$ will be initialized on lines \ref{initwait}, \ref{initwait1} and the termination of the algorithm depends on proving the termination of the loop on line \ref{while}. Now $W(\sigma)$  is intialized by loading the block numbers of the split sets on line \ref{initwait1}. There are only two possiblities after any execution of the loop. Let $W_{m}(\sigma)$ and $W_{m+1}(\sigma)$ represent the state of the variable $W(\sigma)$ before and after one execution of the loop respectively at any given time.  Then either $W_{m}(\sigma) = W_{m+1}(\sigma)\cup\{i\}$
and no splitting has taken place and \emph{i} is the deleted block number, or $W_{m}(\sigma)\cup\{j\}=W_{m+1}(\sigma)\cup\{i\}$ or $W_{m}(\sigma)\cup\{k\}=W_{m+1}(\sigma)\cup\{i\}$
where j and k represent the split blocks and one of them goes into $W_{m}(\sigma)$  if it has fewer incoming transitions. In either case $W_{m}(\sigma)>W_{m+1}(\sigma)$ by Definition \ref{def:definition3}. Therefore $W(\sigma)$ strictly decreases with each iteration of the loop
on line \ref{while}. Since the ordering $\leq$ is well-founded, Algorithm \ref{KMN} must terminate. 
\endproof

Now we only need to show that when Algorithm \ref{KMN} has terminated, it returns 
the Nerode congruence $\equiv$  on states.

\startdef
\noindent {\bf 6.2.3. Proposition.}
{\sl
Let $P_{i}$ be the partition (block set) on the $ith$ iteration of Algorithm \ref{KMN}. For any blocks $B_{j}, B_{k} \in P_{i}$ and any states $p\in B_{j}, q\in B_{k}$  if $j\neq k$ then
$p \not\equiv q$.
}

\noindent {\bf Proof.} 
By induction on the number  $i$ of times the loop on line \ref{while} is executed. 
\vskip 6pt
\noindent\textbf{Basis:} Suppose $i=0$ then clearly the result holds because each block created at line \ref{partition} is distinguishable by the empty string $\epsilon$. 
\vskip 6pt
 \noindent\textbf{Induction Step:} Suppose $i=m>0$. Let us assume that the proposition holds after $m$ executions of the loop.

Consider any $B_{j},B_{k}\in P_{m}$.  During the $m+1$th execution of the loop on line \ref{while} either block $B_{j}$ is split into $B_{j}^{\prime}$ and $B_{j}^{\prime\prime}$  or  $B_{k}$ is split into $B_{k}^{\prime}$ and $B_{k}^{\prime\prime}$  but not both during one execution of the loop (due to line \ref{block}). 

Consider the case when $B_{j}$ is split then for any $p \in B_{j}$,  either $p \in B_{j}^{\prime}$ or $p \in B_{j}^{\prime\prime}$.  But for any $p \in B_{j}$ and $q \in B_{k}$, $p \not\equiv q$ by the induction hypothesis. Therefore, for $p \in B_{j}^{\prime}$  or $p\in B_{j}^{\prime\prime}$  $p \not\equiv q$. Hence the proposition is true for $m+1$th execution of the loop in this case. 

 By symmetry the same argument holds when $B_{k}$ is split.
\endproof

The following Lemma gives a simple, but very effective way to understand Algorithm \ref{KMN}.
Note that this analysis is more like a temporal logic argument than a loop invariant approach.
This approach reflects the non-determinism inherent in the algorithm.

\startdef
\noindent {\bf 6.2.4. Lemma.}
{\sl
\label{lm}
For any states $p , q \in Q$, if $p \not\equiv q$ and initially $p$ and $q$ are in the same block 
$p , q \in B_{i_0}$ then eventually $p$ and $q$ are split into different blocks, 
$p \in B_j$ and $q \in B_k$ for $j \not= k$.
}

\noindent {\bf Proof.} 
Suppose that $p \not\equiv q$ and that initially $p , q \in B_{i_0}$ for some block $B_{i_0}$. 
Since $p \not\equiv q$ then for some $n \geq 0$, and 
$\sigma_1 , \ldots , \sigma_n \in \Sigma$, $$
\lambda^* ( p , \sigma_1 , \ldots , \sigma_n ) \not= 
\lambda^* ( q , \sigma_1 , \ldots , \sigma_n ) .
$$
We prove the result by induction on $n$.
\vskip 6pt
\noindent\textbf{Basis} Suppose $n = 0$, so that $\lambda ( p  ) \not= \lambda ( q  )$. By line  \ref{partition}, 
$p \in B_p$ and $q \in B_q$ and $B_p \not= B_q$. So the implication holds vacuously.
\vskip 6pt
\noindent\textbf{Induction Step} Suppose  $n > 0$ and for some $\sigma_1 , \ldots , \sigma_n \in \Sigma$,
$$
\lambda^* ( p , \sigma_1 , \ldots , \sigma_n ) \not= 
\lambda^* ( q , \sigma_1 , \ldots , \sigma_n ) .
$$
\noindent{\bf(a)} Suppose initially $\delta(p, \sigma_{1})\in B(\sigma_{1}, \alpha)$ and $\delta(q, \sigma_{1}) \in B(\sigma_{1}, \beta)$ for $\alpha \neq \beta$.

Consider when $\sigma = \sigma_{1}$ on the first iteration of the loop on line \ref{while}. Clearly, $B(\sigma_{1}, \alpha ), B(\sigma_{1}, \beta ) \in W(\sigma)$ at this point. Choosing $i= \alpha$ and $j=i_{0}$ on this iteration then since $\delta(p,\sigma_{1})\in B(\sigma_{1},\alpha)$ we have

$$
B_{i_{0}}^{\prime}=\{t\in B_{i_{0}}\;\vert\;\delta(t,\sigma_{1})\in B(\sigma_{1},\alpha)\}\subset B_{i_{0}}
$$
This holds because $q \in B_{i_0}$ but  $\delta( q , \sigma_1 ) \in B( \sigma_1 , \beta )$ and 
$B( \sigma_1 , \alpha ) \not= B( \sigma_1 , \beta )$ so 
$B( \sigma_1 , \alpha ) \cap B( \sigma_1 , \beta ) = \emptyset$ and hence 
$q \not\in B'_{i_0}$.
Therefore $p$ and $q$ are split into different blocks on the first iteration so that 
$p \in B'_{i_0}$ and $q \in B_{i_0} - B'_{i_0}$.

By symmetry,  choosing $i = \beta$ and $j = {i_0}$ then $p$ and $q$ are split on the first loop iteration with 
$q \in B'_{i_0}$ and $p \in B_{i_0} - B'_{i_0}$.

\noindent{\bf (b)} Suppose initially 
$\delta( p , \sigma_1 ) , \delta( q , \sigma_1 ) \in B( \sigma_1 , \alpha )$
for some $\alpha$. Now

$$
\lambda^* (\; \delta( p , \sigma_1 ) , \sigma_2 , \ldots , \sigma_n \; ) \not= 
\lambda^* ( \;\delta( q , \sigma_1 ) , \sigma_2 , \ldots , \sigma_n  \;) .
$$

So by the induction hypothesis, eventually $\delta( p , \sigma_1 )$ and $\delta( q , \sigma_1 )$
are split into different blocks, $\delta( p , \sigma_1 )\in B_{\alpha}$ and 
$\delta( p , \sigma_1 )\in B_{\beta}$. At that time one of $B_{\alpha}$ or $B_{\beta}$
is placed in a waiting set $W(\sigma)$. Then either on the same iteration of the loop on line \ref{while} or on the 
next iteration, we can apply the argument of part (a) again to show that $p$ and $q$ are split into 
different blocks.
\endproof

Observe that only one split block is loaded into $W(\sigma)$
on lines \ref{if}-\ref{endif}.  From the proof 
of Lemma \ref{lm} we can see that it does not matter logically which of these two blocks we insert into $W(\sigma)$. However, 
by  choosing the subset with fewest incoming transitions we can obtain a worst case time complexity 
of order  $O( \vert  \Sigma \vert . n \smsp log_2 \smsp n )$, as we will show.

\startdef
\noindent {\bf 6.2.5. Corollary.}
{\sl
For any states $p , q \in Q$, if $p \not\equiv q$ 
then $p$ and $q$ are in different blocks when the algorithm terminates.
}

\noindent {\bf Proof.} 
Assume that $p \not\equiv q$. 

\noindent{\bf (a)} Suppose at line 3 that $n = \vert Q  \vert$. Then initially, all blocks $B_i$ are singleton sets and 
so trivially  $p$ and $q$ are in different blocks when the algorithm terminates.
\vskip 0.12pt

\noindent{\bf (b)} Suppose at line 3 that $n < \vert Q  \vert$. 

\noindent{\bf (b.i)} Suppose that  $p$ and $q$ are in different blocks initially. 
Since blocks are never merged then the result holds. 

\noindent{\bf (b.ii)} Suppose that  $p$ and $q$ are in the same block initially. 
Since $p \not\equiv q$ then the result follows by Lemma \ref{lm}.
\endproof

We conclude this section by  verifying that our generalisation of Hopcroft's minimisation algorithm does not actually change
its time complexity.

\startdef
\noindent {\bf 6.2.6. Proposition.}
{\sl
If ${\Aut}$ has $n$ states then Algorithm \ref{KMN}
has worst case time complexity $O( \vert  \Sigma \vert . n \log_{2} n )$.
}

\noindent {\bf Proof.} 
Creating the initial block partition on line \ref{partition} requires at most $O(n)$ assignments. 
The block subpartitioning in the loop on line \ref{subpartition} requires at most $O(kn)$ moves of states. Also the 
the initialisation of the waiting lists $W(\sigma)$ in the loop on line \ref{initwait}  requires at most $O(kn)$ assignments.

Consider one execution of the body of the loop starting on line \ref{while}, i.e. lines \ref{bodystart} - \ref{bodyend}. Consider any states 
$p \c q \in Q$ and suppose that $\delta ( p \c  \sigma ) = q$ for some $\sigma \in \Sigma$. Then the state $p$ 
can be: (i) moved into $B'_j$ (line \ref{newBj}), (ii) removed from $B_j$ (line \ref{block}), or (iii) moved into 
$B( \sigma \c  i)$ or $B( \sigma \c  count)$ (lines \ref{newacount}, \ref{newaj}) if, and only if, a block $i$ is being 
removed from $W(\sigma)$
such that $q \in B( \sigma \c  i)$ at that time. 
(Such a block sub-partition $B( \sigma \c  i)$ can be termed a {\it splitter} of $q$.)

Now each time a block $i$ containing $q$ is removed from $W(\sigma)$ its size is less than half of the size when it 
was originally entered into $W(\sigma)$, by lines \ref{if}-\ref{endif}. 
So $i$ can be removed from $W(\sigma)$ at most $O( log_2 \smsp n )$ times. 
Since there are at most $\vert  \Sigma \vert$ values of $\sigma$ and $n$ values of $p$, then the total number of state moves 
between blocks and block sub-partitions is at most $O(  \vert  \Sigma \vert . n \smsp log_2 \smsp n )$.
\endproof

\section{Heuristic Estimation of IKL Convergence}
When the IKL learning algorithm is applied to the problem of learning based testing of software, the question naturally arises, when should we stop testing? 
When the system under test (SUT) is sufficiently small, exhaustive testing can be achieved if we continue until the IKL algorithm converges. 
But how can we detect convergence? 

Traditionally, in automata learning theory, this question is answered by executing an {\it equivalence oracle} 
on the SUT and the hypothesis automaton such as \cite{Norton}. For a DFA learning algorithm such as L* \cite{Angluin87}, learning is continued if the 
equivalence oracle can return a string that is incorrectly learned by the hypothesis DFA, otherwise learning is terminated. However, in the context of black-box
testing a  {\it glass box equivalence oracle}, based on direct comparison of the SUT and the hypothesis automaton, is not acceptable for two reasons: 
\vskip 4pt
\noindent (1) the principles of black-box testing do not allow us to expose the SUT for glass box equivalence checking, and
\vskip 4pt
\noindent (2) even if we ignore (1), in practise there are no glass box equivalence checkers that can compare an arbitrary piece of software (the SUT implementation) with the hypothesis automaton for equivalence. 
\vskip 4pt
Of course, a glass box equivalence oracle can be stochastically approximated by a black-box equivalence oracle based on random queries. 
Random queries are even necessary during LBT when no counterexamples can be found by model checking. However, a purely stochastic
solution to equivalence checking is not possible, as we will discuss below.
Therefore problems (1) and (2) force us to consider other black-box heuristics for estimating convergence of the IKL algorithm. 

\begin{figure}
\begin{centering}
\includegraphics[width=1\columnwidth,height=0.30\paperheight]{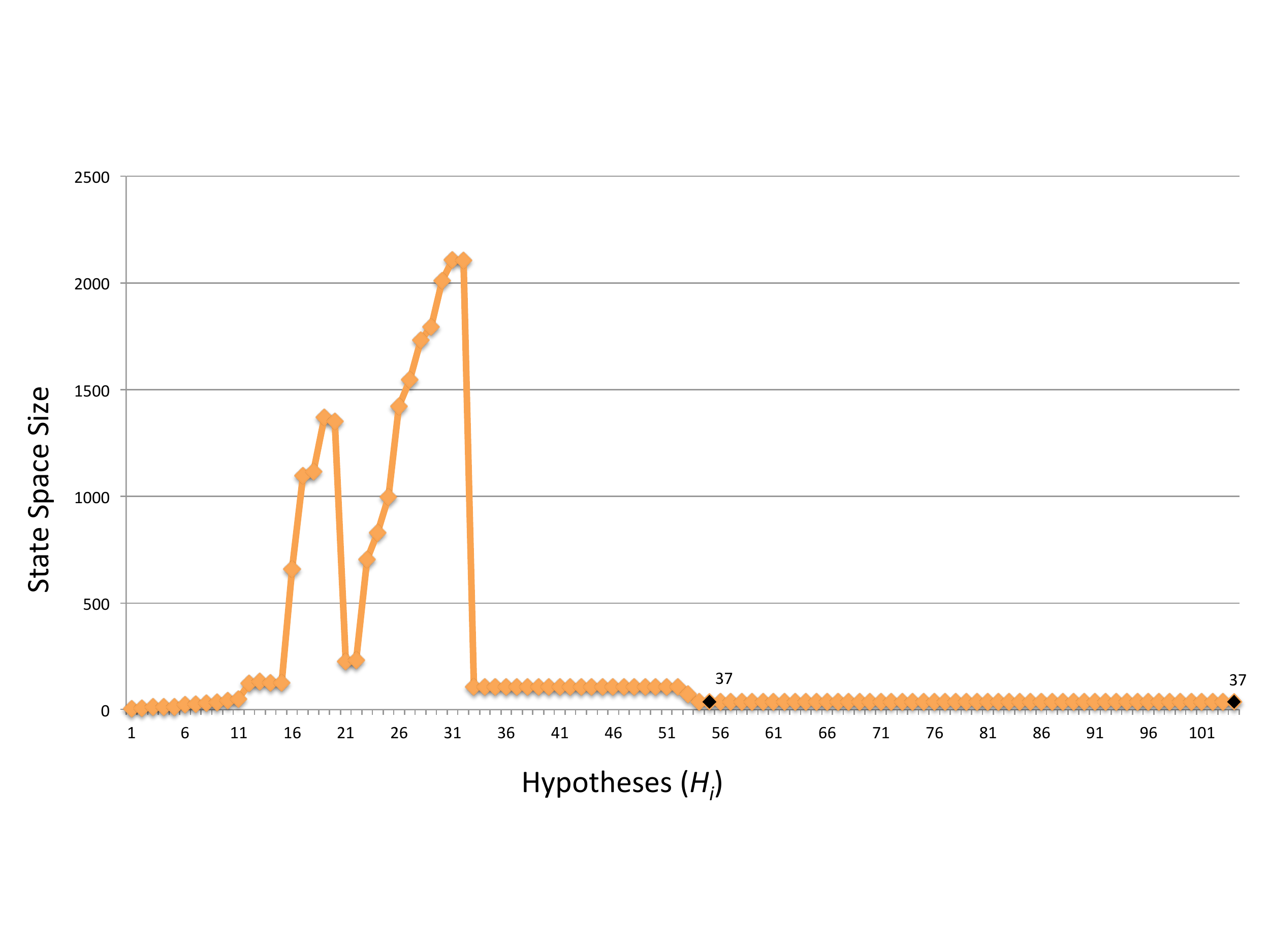}
\par\end{centering}
\caption{\label{fig:spectest} Graph for True and Estimated Convergence for Elevator}
\end{figure}

Figure \ref{fig:spectest} depicts the state space size of successive hypothesis automata $H_i$ ($i = 1, ... 104$) generated by the IKL algorithm while learning and testing a small reactive system against a simple temporal logic specification. In this controlled experiment the SUT was a simplified model of an elevator, with a state space size of 38 states and an input alphabet of 4 symbols. This model is well within the scope of complete learning using IKL, which converges quickly. 

It is natural to consider whether any features of a graph such as Figure  \ref{fig:spectest} can be used to estimate the point of convergence.
This graph is comparable in its structure for all similar experiments that were conducted. 
It shows a succession of peaks, each one well above the state space size of the 
underlying SUT. However at some point these peaks die out and a steady state space size is reached. Each peak and trough seem to indicate a distinct new 
phase in learning, and therefore they do shed some light on the learning activity. However, they clearly do not indicate convergence, which first 
appears in hypothesis automaton $H_{55}$. (In controlled experiments we can apply a glass box equivalence checker to accurately determine convergence.) 

Although we cannot apply glass box equivalence checking between the SUT and  hypothesis automata $H_i$, we can apply it to pairs of
successive hypothesis automata $H_{i}$ and  $H_{i-1}$, since the representations of these are known and visible. We can even iterate this test across $n$ successive hypothesis automata $H_{i} , ... , H_{i-n}$ (by conjunction of the outcomes) which we term {\it $n$-equivalence checking}. After convergence has been achieved,  $n$-equivalence checking will be positive for every value of $n$. 
This gives a heuristic for black box equivalence checking that is more complex than stochastic equivalence checking, since the queries used to generate successive hypothesis automata are not always random. Many arise from model checking counterexamples. It is difficult to say that  queries generated by model checking are randomised, since they are always counterexamples to a specific temporal logic formula, which can strongly bias their structure.

We therefore decided to empirically evaluate the reliability of $n$-equivalence checking as a heuristic indicator of convergence. 
For this evaluation we considered different SUTs with different state space sizes, different temporal logic formulas, and different values of $n$.

We chose two different SUTs, which were models of a simple cruise controller and a simple elevator. The cruise controller model was an 8-state 5-bit Kripke 
structure with an input alphabet size of 5. The elevator model was a 38-state 8-bit Kripke structure with an input alphabet size 
of 4. We considered four different temporal logic test requirements for the cruise controller and six for the elevator. These gave a total of ten convergence 
experiments for the two SUTs. 

Each of these ten experiments was then used to evaluate the $n$-equivalence heuristic for $n = 1, 2, 10, 50$. For $n = 1, 2$ the heuristic completely failed to identify convergence (i.e. the indicator always triggered too early) 
for all ten experiments. For $n = 10$, just two experiments with the cruise controller (the smaller case study) correctly
identified convergence , while eight still failed. Using $n = 50$ all experiments 
correctly identified convergence. However, note that for increasingly large values of $n$ we tended to overestimate the convergence point by an
increasing margin. Table \ref{convresults} summarises the relationship between true convergence and estimated convergence 
for $n = 50$.

\begin{table}
\begin{centering}
\begin{tabular}{|p{2cm}|p{2cm}|p{2cm}|}

\hline 
Requirement & True Convergence & Estimated Convergence\tabularnewline
\hline 
\hline 
1 & $H_{55}$ & $H_{104}$ \tabularnewline
\hline 
2 & $H_{16}$ & $H_{66}$ \tabularnewline
\hline 
3 & $H_{44}$ & $H_{94}$ \tabularnewline
\hline 
4 & $H_{32}$ & $H_{82}$ \tabularnewline
\hline 
5 & $H_{18}$ & $H_{68}$ \tabularnewline
\hline 
6 & $H_{25}$ & $H_{66}$ \tabularnewline
\hline 
\end{tabular}
\medskip
\caption{\label{convresults}True and Estimated Convergence}
\par\end{centering}

\end{table}

These simple experiments suggest that for sufficiently large $n$,  $n$-equivalence, can be used as a reliable heuristic
indicator for convergence. However, further empirical and theoretical analysis still seems necessary to predict the smallest reliable value of $n$ which
minimises the problem of overestimation.

\section{Conclusions}
We have defined and analysed a learning algorithm IKL for deterministic Kripke structures which is efficient for applications in software testing. 
This algorithm extends active incremental learning with new features such as lazy learning and projection. 
We have formally proved the correctness of the IKL algorithm and its main components. We have also empirically evaluated a black box heuristic for detecting convergence of learning, which can be used to terminate testing for small systems under test. 

Incremental learning and projection
combine to make IKL scalable to larger systems under test. Also, incremental and lazy learning combine 
to support frequent generation of hypothesis automata with which we can discover SUT errors much faster than random testing by model checking.
These claims have been empirically evaluated and supported 
in \cite{MeinkeSindhu11} and \cite{Wong2015}.
The IKL algorithm has been implemented in the LBTest tool \cite{MeinkeSindhu13} for learning based testing of 
reactive systems.   

We believe that the efficiency of learning-based testing can be even further improved by
more research on model inference. 
For example, the modular architecture of the IKL algorithm can support experiment with other incremental DFA learning algorithms  instead of 
the ID learning algorithm of Section 4, (e.g.
RPNI2 \cite{Dupont96}). The impact of the frequency of hypothesis automata generation on testing efficiency could then be further 
investigated. When hypothesis generation is very frequent the overhead of model checking is high, and this overhead can slow down the entire 
LBT process. However, if generation is very infrequent, then little use is made of the model checker  to conduct a 
directed search for SUT errors using queries that can falsify the user requirements. This is also inefficient. (Recall the discussion of Section 1.2.) More generally, we could consider an optimal tuning of
the rate of hypothesis automata generation, e.g. based on the estimated density of SUT errors.

The relationship between computational learning and software testing has been a fruitful line of research ever since Weyuker's thesis \cite{Weyuker1983}.
Many fundamental questions remain within the context of learning-based testing. 
For example, the execution of any automata learning algorithm can always be associated with a {\it prefix tree construction}  (see e.g. \cite{Higuera})
based on the query set used.
How can we influence the choice between breadth-first and depth-first search 
for SUT errors using this prefix tree? 
Another important question is whether we 
can find other techniques to generate 
active learner queries besides congruence construction? Such techniques should be aimed at reducing the need for random queries, which can be very inefficient in practise.

We gratefully acknowledge financial support for this research from the Swedish Research Council (VR), the Higher Education Commission (HEC) of Pakistan, 
 and the European Union under project HATS FP7-231620.

%
%

\bibliographystyle{plain}	
\bibliography{refs}		

\end{document}